\documentclass[aps,prd,showpacs,twocolumn,groupedaddress,sort&compress]{revtex4-1}

\usepackage{graphicx}
\usepackage{amsmath}

\begin{document}

\title{Spectra of heavy quarkonia in a Bethe-Salpeter-equation approach}


\author{T. Hilger}
\email[]{thomas.hilger@uni-graz.at}
\affiliation{University of Graz, Institute of Physics, NAWI Graz, A-8010 Graz, Austria}

\author{C. Popovici}
\email[]{carina.popovici@uni-graz.at}
\affiliation{University of Graz, Institute of Physics, NAWI Graz, A-8010 Graz, Austria}

\author{M. G\'{o}mez-Rocha}
\email[]{maria.gomez-rocha@uni-graz.at}
\affiliation{University of Graz, Institute of Physics, NAWI Graz, A-8010 Graz, Austria}

\author{A. Krassnigg}
\email[]{andreas.krassnigg@uni-graz.at}
\affiliation{University of Graz, Institute of Physics, NAWI Graz, A-8010 Graz, Austria}

\date{\today}

\begin{abstract}
In a covariant Bethe-Salpeter-equation approach and with a
rainbow-ladder truncated model of QCD, we investigate the use of an
effective interaction with the goal of reproducing QCD phenomenology. We extend previous
studies and present results for ground and excited meson states in the bottomonium 
and charmonium systems, where the results are surprisingly good for most states. In addition,
we formulate a critical outlook on states with exotic quantum numbers as well as
the light-quark domain.

\end{abstract}


\maketitle

\section{Introduction\label{sec:intro}}


One of the challenges of modern standard-model particle physics is the
description of mesons and baryons via the fundamental
degrees of freedom in quantum chromodynamics (QCD), quarks and gluons. 
The strong-interaction sector of the standard model is beautifully accessible
via the asymptotic freedom of QCD \cite{Gross:1973id,Gross:1973ju,Politzer:1973fx},
but the low-energy properties of hadrons and most prominently confinement and
dynamical chiral symmetry breaking (D$\chi$SB) are accessible from the 
underlying quantum field theory (QFT) only by nonperturbative methods;
in addition, a thorough understanding of these phenomena is paramount
for theoretical hadron physics \cite{Brambilla:2014jmp}.

The recent renaissance of hadron spectroscopy, in particular, is due to the 
fact that this field still offers immediate and influential open question,
e.\,g., the existence, properties, and abundance of hadron states with 
exotic quantum numbers. Any modern comprehensive approach to hadron 
spectroscopy must therefore go beyond the conventional states described
by the quark model---in the meson sector by a standard quark-antiquark ($q\bar{q}$)
configuration---and address these open problems.

Modern approaches to hadron spectroscopy make use of lattice-regularized QCD 
techniques on one hand \cite{Dudek:2007wv,Liu:2012ze,Thomas:2014dpa}, 
and continuum QFT methods on the other \cite{Lucha:2014xla,Wambach:2014vta,Pawlowski:2005xe,Brodsky:2014yha} 
(always see also references therein). Our method of choice in the present work
is the coupled Dyson-Schwinger--Bethe-Salpeter-equation (DSBSE) framework, which
has been successfully applied not only to QCD but also to other strongly coupled
theories, such as $\textrm{QED}_{3}$ or graphene; see for example
\cite{Roberts:2012sv,Bashir:2012fs,Popovici:2013fya} for recent
reviews. 

The DSBSE studies of the past decades have been undertaken at varying levels of sophistication. 
Only in a few particular cases analytical solutions are accessible, such as the
limit of heavy quark mass, where the system can be described by a
variant of heavy-quark effective theory \cite{Popovici:2010mb}, or if
only the IR behavior of the theory is considered \cite{Fischer:2008uz}. All other studies,
and ours as well, rely on truncations that enable numerical investigations. The
infinite tower of coupled DSEs is truncated by restricting the number of equations that are solved
self-consistently, and by compensating for the remaining equations through
sound Ans\"atze for the corresponding Green functions that are not taken into account explicitly.

In particular, we use a basic but symmetry-preserving truncation
to study mesons by solving the quark Dyson-Schwinger equation (DSE) coupled
to the meson $q\bar{q}$ Bethe-Salpeter equation (BSE). Baryon studies are not performed
in the present work, but such studies can be carried out on an equally consistent
footing using covariant quark-diquark or three-quark-equation setups, see e.\,g., 
\cite{Bloch:2003vn,Eichmann:2007nn,Eichmann:2009qa,SanchisAlepuz:2011jn,Eichmann:2012mp,Segovia:2013uga,SanchisAlepuz:2013iia,Sanchis-Alepuz:2014sca} 
and references therein for details.

Despite the difficulty inherent to nonperturbative methods, there are also
immediate benefits, which present an advantage compared, e.\,g., to quark-model studies. 
An excellent example is the possibility to prove results that are exact in QCD.
Prominently, chiral symmetry and its dynamical breaking, along with 
the corresponding constraints, are manifested via the axial-vector Ward-Takahashi 
identity (AVWTI), which serves as a guide for the construction of consistent
corresponding integration-equation kernels \cite{Munczek:1994zz,Bender:1996bb,Bhagwat:2004hn}. 
Furthermore, the AVWTI provides insight on the properties of pseudoscalar
mesons, which in the chiral limit reduces to the well-known Gell-Mann--Oakes--Renner
relation, but can be formulated on general grounds. 
In a symmetry-preserving truncation such as the one used herein, these statements remain
valid and can be checked also numerically. More precisely, our numerical studies
of the pion and its radial excitations show the behavior that is exact in QCD in the chiral limit, 
namely a massless pion ground state with a finite decay constant and massive radially excited
pion states with an exactly zero decay constant each \cite{Maris:1997hd,Holl:2004fr}.
A similar situation is found with respect to electromagnetic properties, where the vector
WTI, also satisfied in RL truncation, and its effects can be tested numerically via charge-conservation
and the behavior of electromagnetic form factors 
\cite{Maris:1999bh,Maris:2005tt,Holl:2005vu,Bhagwat:2006pu,Eichmann:2011vu,Eichmann:2012mp}.

Another important advantage is the manifest covariance of the DSBSE setup, regardless of the truncation
used. It implies immediate usability of both the quark propagators as well as the covariant amplitudes
obtained as solutions of the BSE in any calculation of transition amplitudes between hadrons or
currents and dressed vertex functions. Among the benefits of the covariant four-dimensional setup, 
one also gets direct access to meson states with exotic quantum numbers already at the $q\bar{q}$ level.
As another advantage, we mention the connection to perturbative QCD via the effective interaction
discussed below.

Due to these advantages, the approach has been successfully applied to many individual problems in and 
beyond spectroscopy; concrete examples and therefore 
intrinsically relevant as outlook of this work are, apart from chiral and electromagnetic hadron properties
already cited above,
strong hadron decay widths \cite{Jarecke:2002xd,Mader:2011zf}, valence-quark distributions of pseudoscalar mesons
\cite{Nguyen:2010ph,Tandy:2010dw,Nguyen:2011jy,Chang:2013nia},
studies of tensor mesons \cite{Krassnigg:2010mh,Fischer:2014xha} and extensions of this setup to QCD at finite
temperature \cite{Maris:2000ig,Horvatic:2007qs,Blank:2010bz}.

While all these individual results and studies provide quite a wealth of information and a large portion
was even computed with the same model (which is also used here), there is no comprehensive meson, let alone hadron
study so far, and our work is the first step towards one. At the level of RL truncation, it remains to be shown
what the range of success of such a comprehensive endeavor can be or whether it is possible at all. As a final
part of motivation, it is helpful to mention that even a successful study of radial meson excitations
such as the one presented herein has been generally doubted and deemed impossible, which makes our results
relevant in the first place and remarkable at the same time.

We note at this point that our calculations have been performed using Landau-gauge QCD in
Euclidean momentum space. Progress made using the Minkowski-space formulation of the
BSE to study the $q\bar{q}$ system is ongoing and can be traced via 
\cite{Sauli:2008bn,Carbonell:2010zw,Sauli:2011aa,Frederico:2011ws,Sauli:2012xj,%
Carbonell:2013kwa,Frederico:2013vga,Carbonell:2014dwa,Hall:2014dua}. 
Calculations in the Coulomb gauge of QCD are slightly different in terms
of numerical feasibility as well as the particular systems or domains
that are more easily described. For details, see 
\cite{Alkofer:2005ug,Rocha:2009xq,Popovici:2010mb,LlanesEstrada:2010bs,Cotanch:2010bq,Popovici:2011yz,Popovici:2011wx} 
and references therein.

The paper is organized as follows: in Sec.~\ref{sec:truncation} we
review the benefits and caveats of the rainbow-ladder (RL) truncation of
the DSBSE system. Section \ref{sec:model} contains the details on
the effective interaction used herein. Results and conclusions are
presented in Secs.~\ref{sec:results} and \ref{sec:conclusions},
respectively.

\section{Rainbow-ladder truncation\label{sec:truncation}}

For comprehensive phenomenological modeling with a realistic
effective interaction, the truncation of choice currently is the
RL truncation of the quark-DSE--meson-BSE system.

Studies beyond-RL truncation are often exploratory in nature and use an interaction
simple enough to deal with the complexity of particular aspects of the infinite tower
of the DSEs and the corresponding BSE setup but too simple to retain all features
required for a successful spectroscopy of hadrons, let alone the calculation of transition
matrix elements \cite{Bender:1996bb,Bhagwat:2004hn,Gomez-Rocha:2014vsa}.

In more sophisticated settings, the effective interaction is realistic overall or at least 
in some particular aspect of the diagrammatic setup of the truncation scheme
\cite{Watson:2004jq,Watson:2004kd,Fischer:2005en,Matevosyan:2006bk,Matevosyan:2007cx,Fischer:2008wy,%
Fischer:2009jm,Williams:2009wx,Williams:2014iea,Sanchis-Alepuz:2014wea}. 
As an alternative, other studies have approached the problem of constructing a consistent BSE kernel
for a given quark-gluon vertex on a more general footing, see \cite{Chang:2009zb,Heupel:2014ina} and references therein.
However, such investigations have never been comprehensive due to the numerical and conceptual difficulty involved.
In addition, neither the conceptional problems of the BSE such as the determination of the analytic
structure of the quark propagator or the possible spurious nature of some excited
states, nor the phenomenological problems encountered, e.\,g., in the
description of axial-vector meson states, were satisfactorily resolved.
In this sense, an RL study can be considered reasonable and most notably constructive 
towards the goal of a comprehensive phenomenological application of the DSBSE approach.

It is important to note here again that RL truncation satisfies the relevant (axial-vector and vector) 
Ward-Takahashi identities
(see e.\,g.~\cite{Maskawa:1974vs,Aoki:1990eq,Kugo:1992pr,Bando:1993qy,Munczek:1994zz,Maris:1997hd,Maris:1999bh,Maris:2000sk})
and thus remains true to the underlying QCD in the corresponding respects. A reliable
numerical setup (ours is detailed in \cite{Krassnigg:2008gd,Blank:2010bp,Krassnigg:2010mh,Blank:2011qk}) 
is important, in particular, with increasing quark mass.

\section{Bound state equation and model interaction\label{sec:model}}

We note at this point that meson studies like ours can be conducted equally well using
the homogeneous BSE or an analogous but more general inhomogeneous vertex BSE, see, 
e.\,g.~\cite{Maris:2000ig,Bhagwat:2007rj,Blank:2010sn}.
Herein we employ the homogeneous $q\bar{q}$ BSE in RL truncation which reads 
\begin{eqnarray}\nonumber
\Gamma(p;P)&=&-C_F\!\!\!\int^\Lambda_q\!\!\!\!\mathcal{G}((p-q)^2)\; D_{\mu\nu}^\mathrm{f}(p-q) \;
\gamma_\mu \; \chi(q;P)\;\gamma_\nu \\ \label{eq:bse}
\chi(q;P)&=&S(q_+) \Gamma(q;P) S(q_-)\,,
\end{eqnarray}
where $q$ and $P$ are the quark-antiquark relative and total momenta, respectively, and 
the (anti)quark momenta  are chosen as $q_{\pm} = q\pm P/2$.
This equation requires knowledge of the quark propagator $S(p)$, which is obtained
from its DSE ($C_F=4/3$ is the Casimir color factor)
\begin{eqnarray}\nonumber
S(p)^{-1}  &=&  (i\gamma\cdot p + m_q)+  \Sigma(p)\,,\\\label{eq:dse}
\Sigma(p)&=& C_F\!\!\int^\Lambda_q\!\!\!\! \mathcal{G}((p-q)^2) \; D_{\mu\nu}^\mathrm{f}(p-q)
\;\gamma_\mu \;S(q)\; \gamma_\nu \,.
\end{eqnarray}
In the above, the effective interaction is denoted by
$\mathcal{G}$ and will be specified in detail below.  $\Sigma$ is
the quark self-energy, $m_q$ is the current-quark mass,
$D_{\mu\nu}^\mathrm{f}$ represents the free gluon propagator and $\gamma_\nu$ is
the bare quark-gluon vertex's Dirac structure.  Dirac and flavor indices are omitted for brevity.
$\int^\Lambda_q=\int^\Lambda d^4q/(2\pi)^4$ denotes a
translationally invariant regularization of the integral, with the regularization scale
$\Lambda$ \cite{Maris:1997tm}.

The evolution of the RL effective interaction $\mathcal{G}$ started from a Dirac-$\delta$
in momentum space, which reduces the coupled integral equations to a set of coupled algebraic
equations \cite{Munczek:1983dx}. For several studies on different levels of sophistication
with regard to the numerical treatment of the evaluation of the quark-propagator dressing functions needed
as input in the BSE, additions and modifications were made to this term such as a 2-loop perturbative-QCD
contribution and an Ansatz for the infrared behavior \cite{Munczek:1991jb,Jain:1993qh} as 
well as one-loop perturbative QCD together with modified 
\cite{Richardson:1978bt,Higashijima:1983gx,Aoki:1990eq,Higashijima:1991de,Kugo:1992zg,Yamanaka:2013zoa} 
or additional strength in the intermediate-momentum
regime both with \cite{Maris:1997tm} and without the $\delta$-term \cite{Maris:1999nt,Jarecke:2002xd,Qin:2011xq,Fischer:2014cfa}.
Further modifications include a focus on low and intermediate momenta with less emphasis
on the perturbative part \cite{Frank:1995uk} and even an ultra-violet (UV) finite version \cite{Alkofer:2002bp},
which emphasized the importance of the intermediate-momentum domain for spectroscopy; in
addition, it was shown that the influence of the far-infrared behavior of the interaction
on meson properties is minor \cite{Blank:2010pa} for the concrete case of the $\rho$ meson
mass and decay constant.

In addition, alternative approaches have been proposed, 
where the effective coupling is adjusted via the quark-gluon vertex such that the quark 
mass function remains independent on the normalization point, and the correct
asymptotic behavior is preserved \cite{Fischer:2003rp,Alkofer:2008et}. Furthermore, models 
have been constructed for the effective interaction such that gauged lattice quark propagators 
are reproduced via the quark DSE solutions \cite{Bhagwat:2003vw,Krassnigg:2003dr,Krassnigg:2003wy,Fischer:2005nf,Eichmann:2008zz}. 

We also mention here that corrections to the RL truncation are expected to diminish with
increasing quark mass, which prompted the investigations in \cite{Eichmann:2008ae,Eichmann:2008ef}, where
a quark-mass dependence was introduced in the main interaction parameters and the setup was
tested for both meson and baryon states. Recently, more evidence towards the necessity of a
quark-mass dependence of a phenomenologically successful RL study of meson properties has emerged 
\cite{Williams:2014iea}. In our present work and strategy, we include this possibility in a natural
way, as is detailed below.

The model of Ref.~\cite{Maris:1999nt} was used extensively in the past to investigate the meson spectrum
and various meson properties with great success, in particular in the pseudoscalar and vector 
channels (see also the references given above) and this parameterization is our choice herein as well.
It reads
\begin{equation}
\label{eq:interaction} 
\frac{{\cal G}(s)}{s} =
\frac{4\pi^2 D}{\omega^6} s\;\mathrm{e}^{-s/\omega^2}
+\frac{4\pi\;\gamma_\mathrm{m} \pi\;\mathcal{F}(s) }{1/2 \ln
[\tau\!+\!(1\!+\!s/\Lambda_\mathrm{QCD}^2)^2]}.
\end{equation} 
The first term is characterized by the parameters $\omega$ (which corresponds to an effective
inverse range of the interaction) and $D$ (which acts like an overall strength)
and determines the intermediate-momentum part of the 
interaction, while the second describes the UV and produces the
correct one-loop perturbative QCD limit. 
${\cal F}(s)= [1 - \exp(-s/[4 m_\mathrm{t}^2])]/s$ where $m_\mathrm{t}=0.5$~GeV,
$\tau={\rm e}^2-1$, $N_\mathrm{f}=4$, $\Lambda_\mathrm{QCD}^{N_\mathrm{f}=4}=
0.234\,{\rm GeV}$, and $\gamma_\mathrm{m}=12/(33-2N_\mathrm{f})$, which is left unchanged from Ref.~\cite{Maris:1999nt}.

In addition to the current-quark masses, $\omega$ and $D$ are those
parameters of the interaction whose impact on meson spectroscopy provides the focus of this work.
It was found already in \cite{Maris:1999nt} that pseudoscalar- and vector-meson ground-state
properties remained unchanged for light mesons if one varies $\omega$ in the 
range $[0.3,0.5]$ GeV and determines $D$ by keeping their product fixed to the phenomenologically successful value
of $D\times\omega=0.372$ GeV${}^3$. Essentially, this corresponds to the statement
that \emph{ground} states, which in the quark model have orbital angular momentum $l=0$,
have properties that do not depend strongly on the effective range of the long-range (intermediate-momentum)
piece of the strong effective interaction; this situation was contrasted by the case
of radially excited meson states \cite{Holl:2005vu,Holl:2004un,Qin:2011xq} 
and other types of excitations, most prominently those corresponding to $l\ne 0$ in the quark model
\cite{Krassnigg:2009zh,Krassnigg:2010mh}.
These dependences can be used to sufficiently constrain all parameters of the interaction,
in particular both $\omega$ and $D$. In fact, the more states our model is compared to, the more
difficult it is to achieve a decent overall description, which is a real challenge both for the 
model setup as well as for RL truncation itself.

After a recent quarkonium study (restricted to the $D\times\omega=$ const.\  prescription but
still successful for the ground-states in bottomonium and, to some extent, also charmonium) was
presented in \cite{Blank:2011ha}, herein we require a more comprehensively successful description of experimental data,
in particular including radially excited states in each $J^{PC}$ channel. To attempt such an agreement with 
experiment, we vary $\omega$ and $D$ independently along the lines of a strategy outlined in detail in \cite{Popovici:2014pha},
where this investigation was already carried out for bottomonium. In short, the parameters 
are fitted to a set of representative experimental level splittings first; in a second step,
the quark mass is determined by a least-squares fit to the ground-state bottomonium masses known experimentally.

Here, we add the case of charmonium and discuss the consequences of our results for states with exotic quantum numbers
as well as a number of states found experimentally, whose quantum numbers have not yet been determined completely.
It is noteworthy that we fit the values of $\omega$ and $D$ separately for each current-quark mass, such that
a quark-mass dependence of these parameters will emerge. The next section reviews the situation in 
bottomonium and details our charmonium results.

\section{Results and discussion\label{sec:results}}

\begin{figure*}
\centering
 \includegraphics[width=0.62\textwidth]{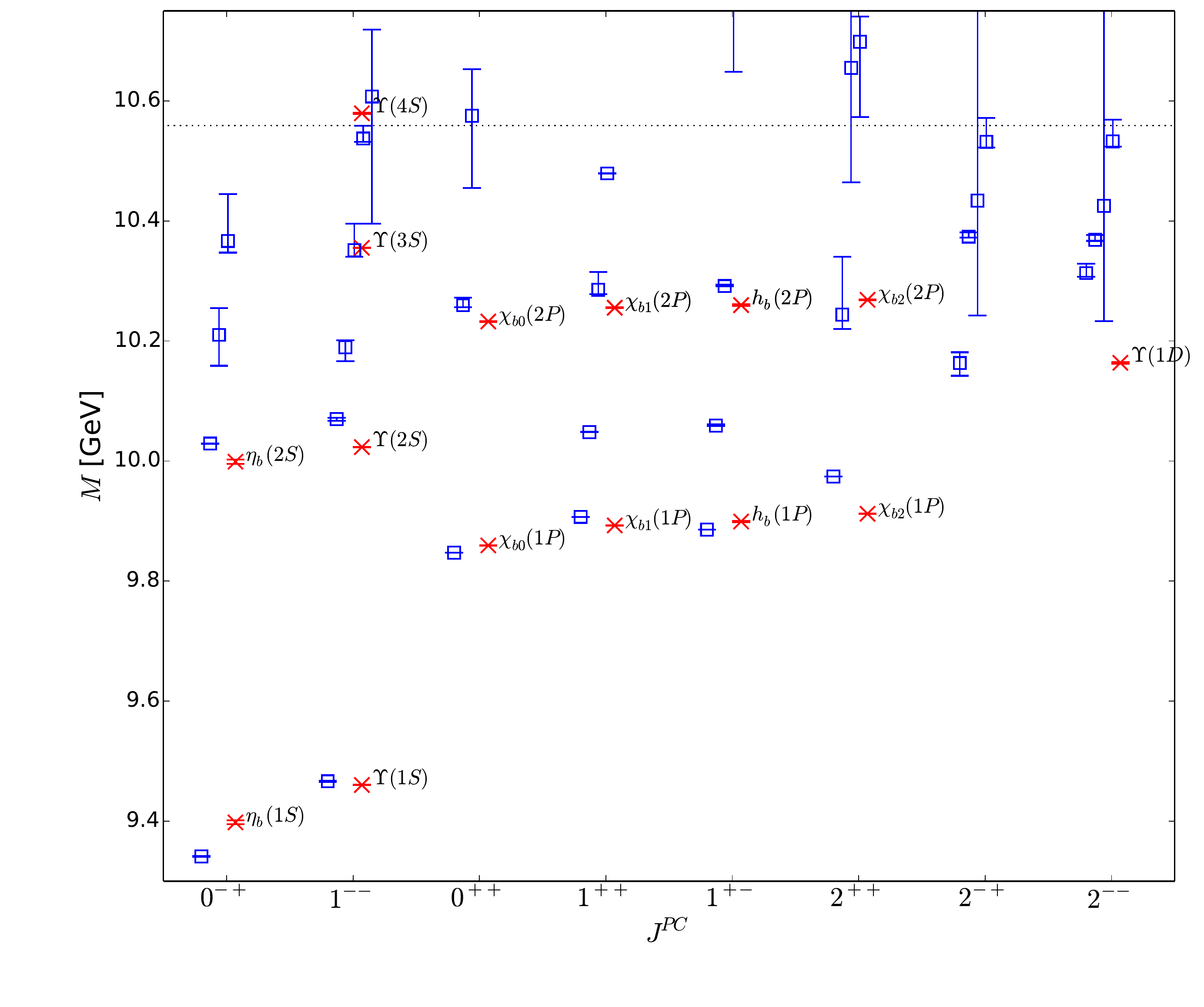}
\caption{(Color online) Bottomonium spectrum: calculated (blue boxes) versus
  experimental data (red crosses) \cite{Olive:2014rpp}. Theoretical error bars represent
  uncertainties from extrapolation techniques, where necessary (see text). The horizontal
  dotted line marks the open-flavor threshold.}
\label{fig:bottomonium}      
\end{figure*}

\begin{figure*}
\centering
 \includegraphics[width=0.62\textwidth]{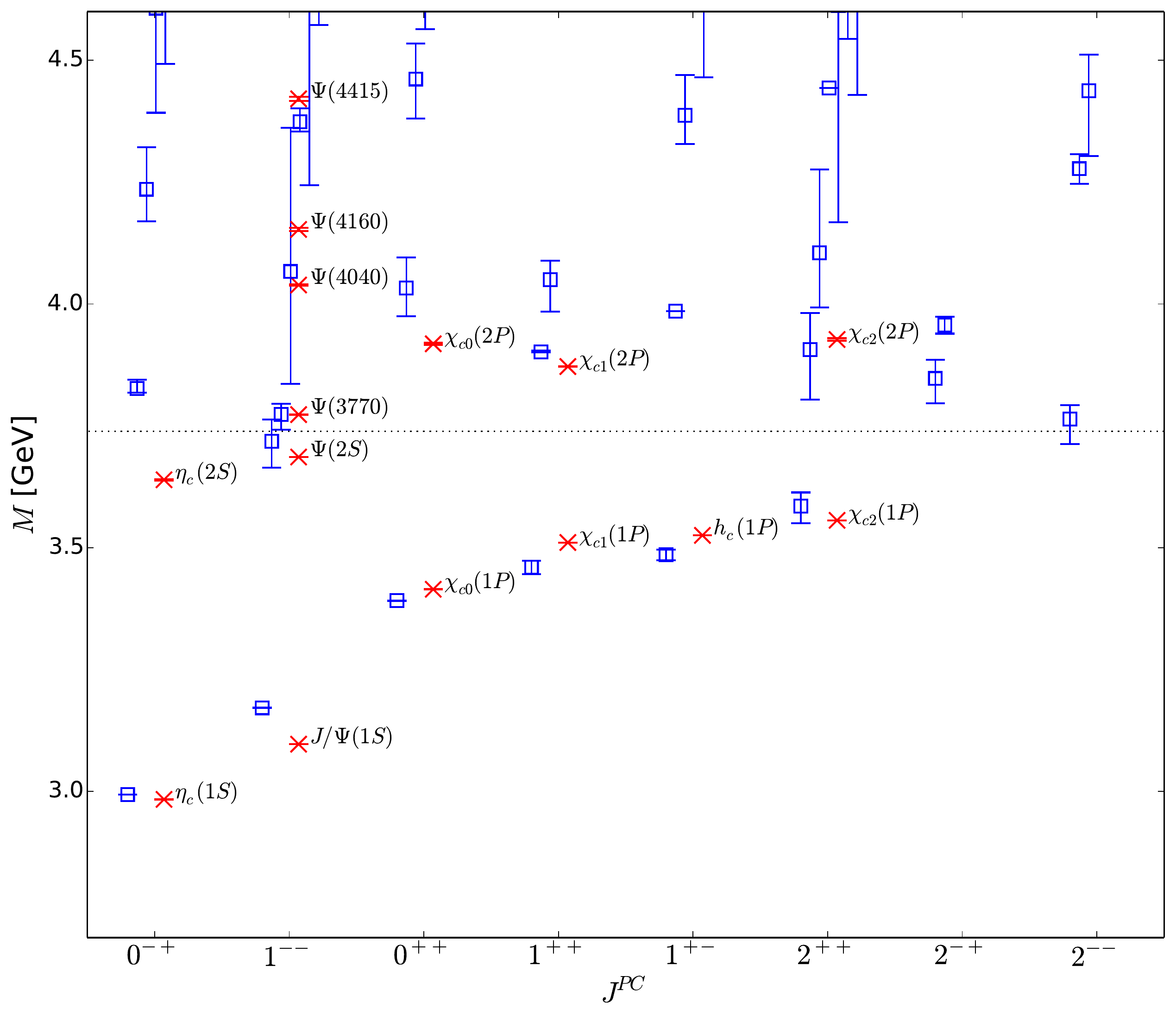}
\caption{(Color online) Charmonium spectrum: calculated (blue boxes) versus
  experimental data (red crosses) \cite{Olive:2014rpp,Brambilla:2010cs}. Theoretical error bars represent
  uncertainties from extrapolation techniques, where necessary (see text). The horizontal
  dotted line marks the open-flavor threshold.}
  \label{fig:charmonium}      
\end{figure*}

In \cite{Popovici:2014pha} we obtained our best fit to the bottomonium spectrum for $m_\mathrm{b}=3.635$ GeV (given at a 
renormalization point $\mu=19$ GeV) together with $\omega=0.7$ GeV and $D=1.3$ GeV${}^2$. The results are
shown as blue boxes in Fig.~\ref{fig:bottomonium}, together with experimental data \cite{Olive:2014rpp}, shown as red crosses.
In the same way, we fitted the charmonium spectrum and obtained $m_\mathrm{c}=0.855$ GeV (given at a 
renormalization point $\mu=19$ GeV) together with $\omega=0.7$ GeV and $D=0.5$ GeV${}^2$; the results are shown 
in Fig.~\ref{fig:charmonium}, together with experimental data \cite{Olive:2014rpp,Brambilla:2010cs}. 
We note that our error bars, where relevant, come from 
extrapolated results in situations where propagator singularities prohibit a direct cacluation; details on the source of
this problem and our extrapolation strategy can be found in the appendix as well as the appendices of 
\cite{Krassnigg:2010mh,Blank:2011ha,Dorkin:2013rsa}.

In addition, it is important to note that our results correspond to bound states and not resonances due to the
effect of the truncation: open hadronic decay channels are not contained in the RL-BSE interaction kernel. Hadronic
(and other) decay width or properties are computed from the solutions of the BSE as well as the quark-propagators
in a semi-perturbative fashion. In particular, as mentioned in the introduction, efforts have been made towards the
calculation of vector to pseudoscalar-pseudoscalar decays for light and strange mesons \cite{Jarecke:2002xd} 
as well as the $\Delta$ in the baryon sector \cite{Mader:2011zf}. While it is both natural and desirable for our study to include
such results in the future, the effort to achieve them is clearly beyond the scope of the present manuscript.
For the moment, we can only caution the reader when comparing our results to experimental data above the respective
open-flavor thresholds. Apparently, this issue is of lesser importance in the bottomonium case than for charmonium. 
To illustrate this and to facilitate the analysis of our results we have marked the thresholds in our figures by 
horizontal dotted lines.

We begin the discussion with bottomonium shown in Fig.~\ref{fig:bottomonium}, where we find very good agreement 
between our results and well-established experimental data. Most splittings are well reproduced, in particular between ground 
and radially excited states in each channel. It is noteworthy that we find the correct level ordering of 
the first radially excited $0^{-+}$ and $1^{--}$ states in the bottomonium system; in a similar fashion, level orderings 
are well reproduced with a few exceptions. In general there is a clear identification of each ground- and first radially
excited state known experimentally with one of our results. However, there are a few caveats. While a slight mismatch for
the $2^{--}$ ground state is apparent, we expect on the basis of \cite{Blank:2011ha} that this can be cured by further 
fine-tuning of model parameters. Higher radial excitations than the first are mostly unclear at the moment due to both theoretical 
and experimental uncertainties overall, except for the vector channel, where the experimental situation is excellent due
to the prominent coupling to $e^+ e^-$. We find excellent agreement for the $\Upsilon(3S)$, but at the same time one
lower result without an experimental match. Further investigations are needed to clarify the role of this state, and 
are currently on their way. A similar situation is encountered in the axial-vector channels, where one extra calculated result each 
appears in between the ground and first radially excited experimental state. At the present time however, we have no reliable tools 
at hand to to determine, e.\,g., whether or not these extra states might be spurios solutions of the BSE. Consequently, we have 
to defer a more in-depth discussion to a later time. In the meantime, similarly to the vector case, we will use means 
beyond spectroscopy to determine the role of these states and report the results in future publications. 

For charmonium presented in Fig.~\ref{fig:charmonium} the state-identification between experiment and calculation is even better
and much clearer:
no extra calculated states are encountered in the domain of the ground states and first radial excitations. Again, splittings
between radially excited and ground states in each channel are very well reproduced; the same is true for the level orderings
with the exception of the $\eta_c(2S)$, which is too heavy in our study. In addition to this excellent overall agreement, it
is most notable, how closely the radial excitations in the vector channel can be matched, even beyond the $\Psi(2S)$. With regard
to the matching and quality of the description of experimental data in both bottomonium and charmonium, we note again that
our search for the optimal model parameters was carried out within the setup of the particular model chosen. We expect
that better agreement can be reached by further fine-tuning of the shape of the model interaction.

While we defer a detailed discussion of exotic-state masses in the various $J^{PC}$ channels to future works, we give a brief
outlook already at this point: states with exotic quantum numbers are generally low in our RL study compared to expectations
from other approaches. More concretely, we find the $0^{--}$ and $1^{-+}$ to be lowest in both bottomonium and charmonium. 
In the former case, they lie even below the $l=1$ ground states at $\sim 9.7$ GeV, while in the latter they lie
in the same region as the $l=1$ ground states at $\sim 3.6$ GeV.

\begin{figure}
\centering
 \includegraphics[width=0.99\columnwidth]{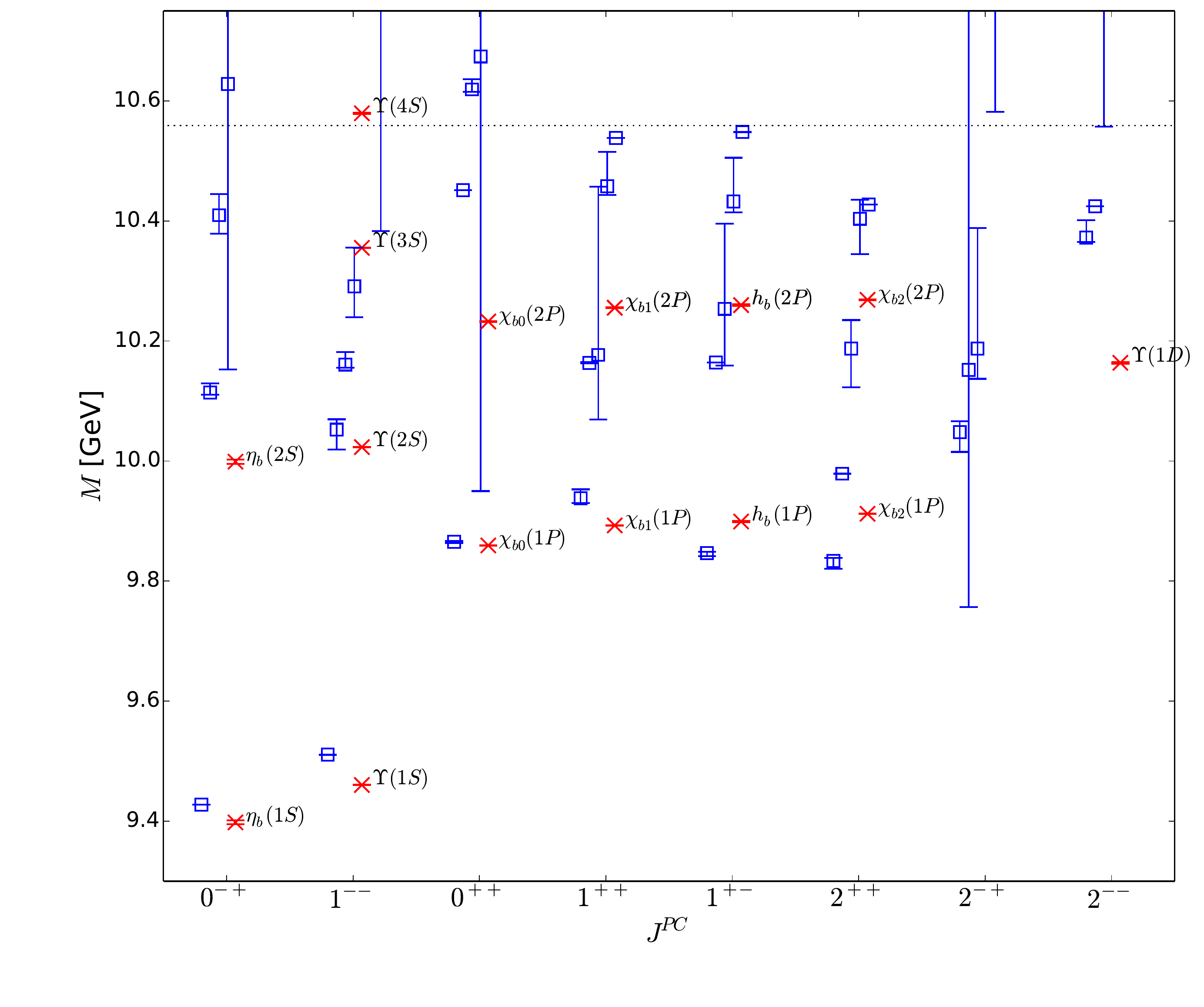}
\caption{(Color online) Bottomonium spectrum cross-check: Same style as Fig.~\ref{fig:bottomonium}, but
computed with the optimal parameters from charmonium (see text).}
\label{fig:bottomonium-cross}      
\end{figure}

\begin{figure}
\centering
 \includegraphics[width=0.99\columnwidth]{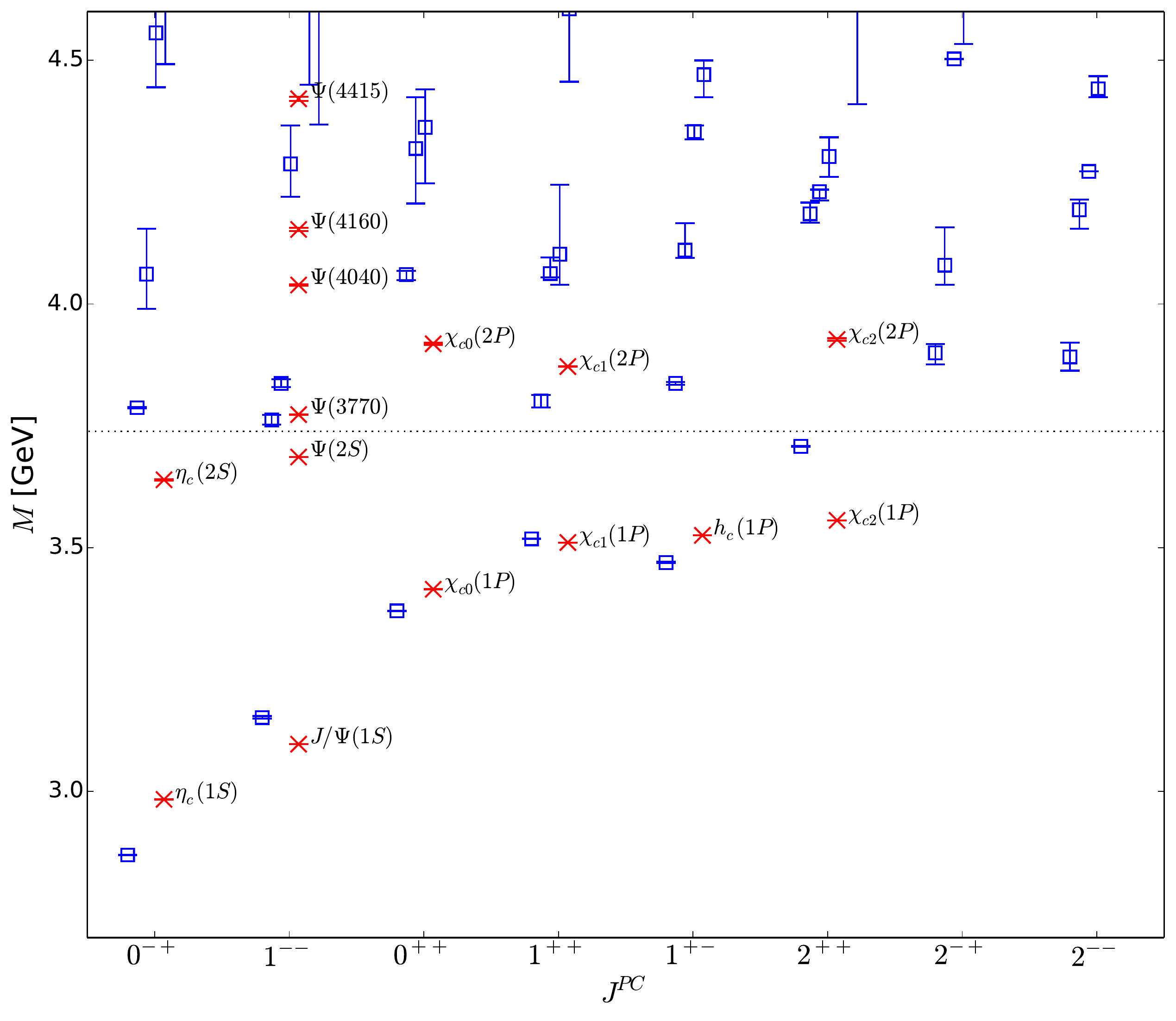}
\caption{(Color online) Charmonium spectrum cross-check: Same style as Fig.~\ref{fig:charmonium}, but
computed with the optimal parameters from bottomonium (see text).}\label{fig:charmonium-cross}      
\end{figure}

Before concluding, we present some evidence as to how feasible a description of both charmonium and bottomonium is
with the same set of model parameters. To illustrate this, we present two more figures analogous to 
Figs.~\ref{fig:bottomonium} and \ref{fig:charmonium}, but with the other set of the model parameters $\omega$ and $D$, 
respectively. Thus, the bottomonium spectrum with $\omega=0.7$ GeV and $D=0.5$ GeV${}^2$ is shown in 
Fig.~\ref{fig:bottomonium-cross} and the charmonium spectrum with $\omega=0.7$ GeV and $D=1.3$ GeV${}^2$ is 
shown in Fig.~\ref{fig:charmonium-cross}. We observe that in both cases the good description of level orderings, radial
splittings, and identification of states is destroyed.

\section{Conclusions and outlook\label{sec:conclusions}}

It has been speculated in the past how successful a comprehensive RL
truncated DSBSE model of hadrons, or at least mesons, can be. The
negligence of all but the $\gamma_\mu$ component of the quark-gluon
vertex was generally believed to be too drastic as to still allow for
a reasonable phenomenological description of the various splittings in
the meson spectrum. These expectations were based on caclulations done
mainly with light quarks and on a domain of the model parameters restricted by anchoring
them in the light quark domain. In particular, primary anchors were the pion mass and 
decay constant as well as the chiral condensate. However, in this respect it is very important to 
differentiate between the light- and heavy-quark cases, since corrections to RL
truncation are expected to diminish with increasing quark mass. Therefore, only an RL study
anchored in the heavy-quark domain can be expected to be well-suited for comprehensive purposes.
In addition, a quark-mass dependence in the effective interaction may provide a better way
towards a unified description of meson spectra over the entire range of experimentally available
quark masses.

In order to fully establish the DSBSE framework as an adequate and valuable complementary alternative
to the quark model and other non-perturbative approaches to QCD, it is imperative to attempt a
comprehensive study of hadrons. As a first step, the requirements for such a study must be taken
beyond a collection of individual results towards a unified model study with as wide a scope as possible.
We have identified an RL truncated DSBSE setup with a sophisticated model interaction as a candidate for
such a study and presented the first step here. 
In our study of heavy quarkonia we have determined the sets of model parameters that optimize 
an RL DSBSE description of the meson spectrum, including both ground states and radial excitations
for the first time. We found good overall agreement with experimental data to a degree well beyond
the general expectations regarding the truncation used herein. Nonetheless, there are caveats, in particular
extra states in the vector ($1^{--}$) and axial-vector ($1^{++}$ and $1^{+-}$) channels in 
bottomonium as well as a lack of clarity in 
the computational outcome for the higher radial excitations, both of which are subject of
ongoing further studies.

The next steps are to extend this study to the light-quark sector, investigate the role of extra states
in the calculated results as well as attempt to identify experimental states with undetermined quantum numbers
or some of the $X$, $Y$, and $Z$ states, respectively, with appropriate results from our calculations. The 
set of results will include masses and decay constants at first, and later also comprise electromagnetic as
well as hadronic width and properties.
We emphasize that this includes to present and discuss concrete results for states with exotic quantum numbers. 
In the course of our studies, we may allow even more free parameters or a different functional form in the 
effective interaction, whose parametric degrees of freedom have not yet been fully exploited, in order to
more effectively fine-tune the results, if necessary.

\begin{acknowledgments}
We acknowledge helpful conversations with M.~Blank, G.~Eichmann, and R.~Williams.
This work was supported by the Austrian Science Fund (FWF) under project no.\ P25121-N27.
\end{acknowledgments}

\appendix

\section{Technicalities\label{sec:tech}}

\begin{figure}
\centering
 \includegraphics[width=0.99\columnwidth]{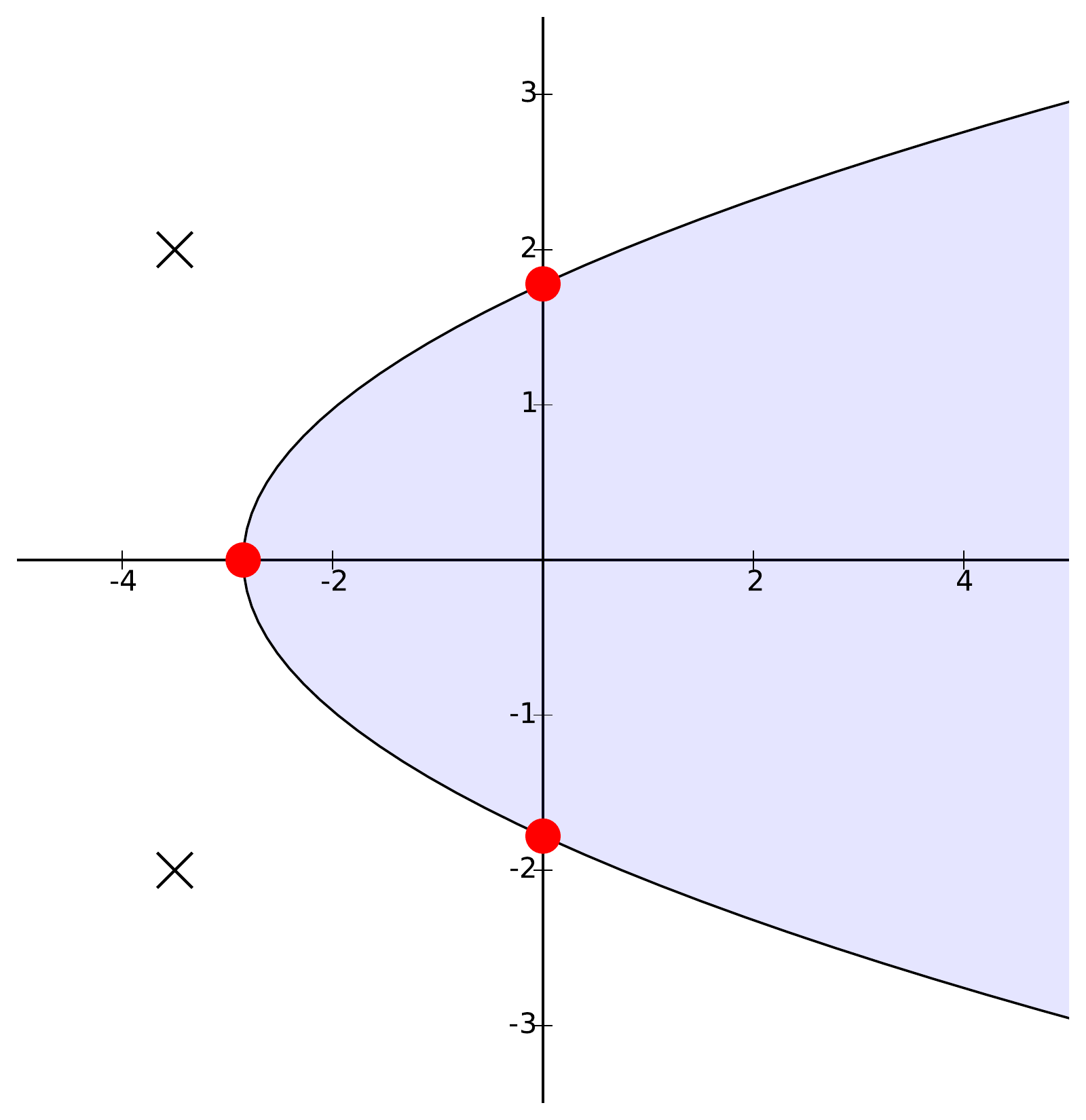}
\caption{(Color online) Integration domain (light blue area) with parabolic boundary in the complex $q_\pm^2$-plane,
on which the (anti)quark propagator dressing functions need to be known numerically. The red dots identify the 
intersection points with the real and imaginary axes; the crosses illustrate the typical location of singularities
in the dressing functions limiting the integration domain (see text).}\label{fig:parabola}      
\end{figure}
In the Euclidean-space formulation of the DSBSE approach to mesons, the BSE
contains two dressed (anti)quark propagators that depend on the momenta $q_\pm$ as given in Eq.~(\ref{eq:bse})
and below. With a timelike total momentum $P$ and the integration momentum $q$ being the gluon momentum, one
needs to compute the propagator dressing functions in a region of the complex $q_\pm^2$-plane that lies inside
a parabolic boundary, stretching towards real positive infinity, indicated as the light blue area in Fig.~\ref{fig:parabola}.
Assuming a real, positive bound-state mass of $M$ with $P^2=-M^2$ and two equal-mass constituents, 
the corresponding integration domain can be defined
\begin{figure}
\centering
 \includegraphics[width=0.79\columnwidth]{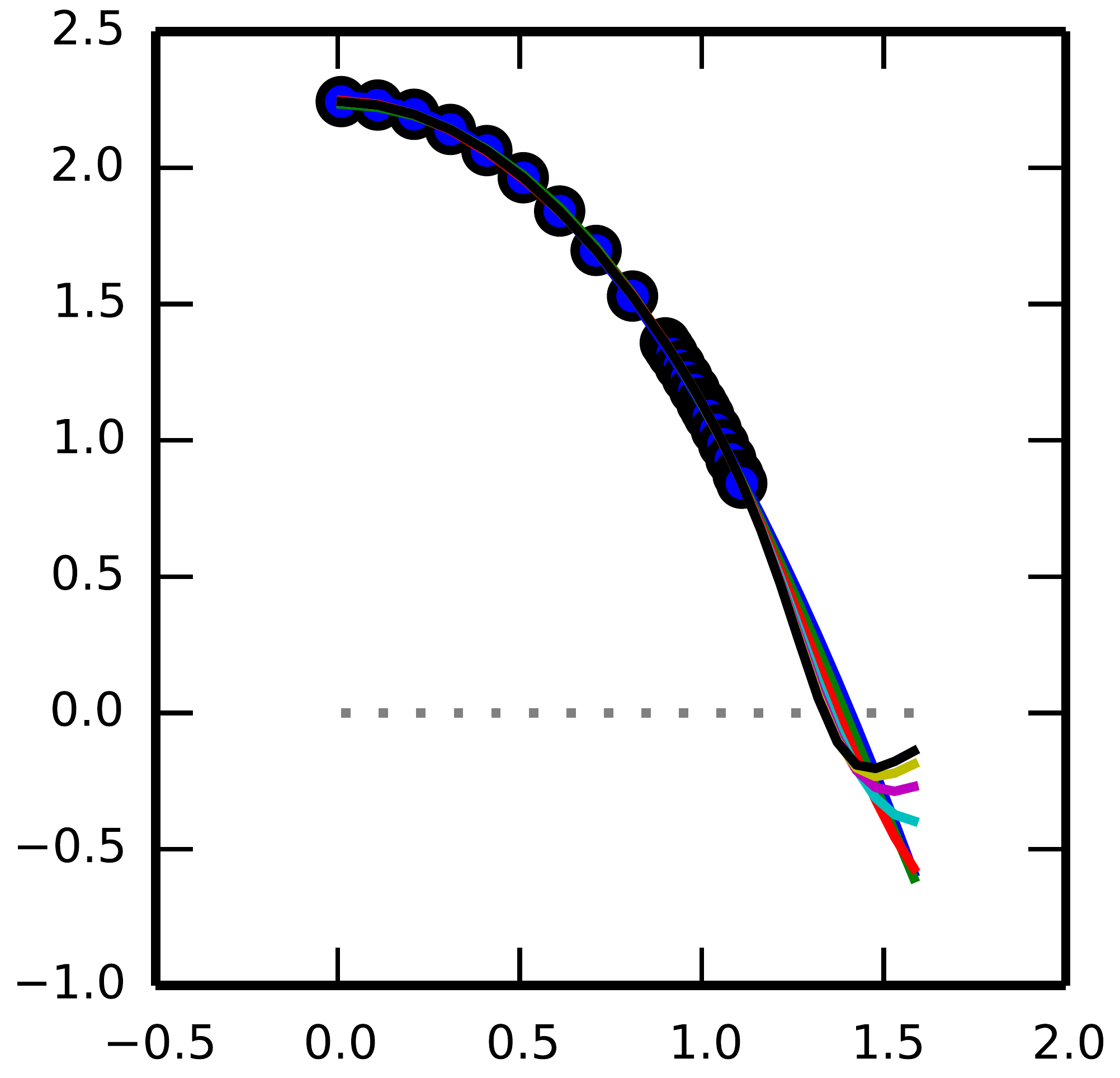}
\caption{(Color online) Example for extrapolation of $\tilde{\lambda}(P^2)$ curves with different $N$
represented by different colors to obtain the bound state mass $M$ where $\tilde{\lambda}(P^2)=0$, as defined
in Eq.~(\ref{eq:lambdapole}) and below, for a pseudoscalar radial excitation.}\label{fig:ext1}      
\end{figure}
via the three intersection points of the parabolic boundary with the real and imaginary axes, at $(-M^2/4,0)$ and 
$(0,\pm M^2/2)$, respectively, marked by the red dots in Fig.~\ref{fig:parabola}. In practice, keeping the numerical 
setup straight-forward \cite{Blank:2010bp}, this means that any singular structure in the propagator dressing functions puts a limit on 
the maximum bound-state mass obtainable via standard methods; a typical scenario is depicted in Fig.~\ref{fig:parabola},
where singularity positions are marked with black crosses. While a ground-state calculation is mostly safe from 
such problems, excited states mostly lie above the mass range obtainable directly. As a simple way to deal with this,
one can resort to extrapolation techniques. First steps had been taken in \cite{Krassnigg:2010mh} and a more sophisticated setup
has been used in \cite{Blank:2011ha} and also herein. As a result, the extrapolation introduces an uncertainty in 
our calculation, which we acknowledge by plotting error bars on our resulting masses. To immediately illustrate 
typical cases, we present extrapolations for a pseudoscalar and a scalar radially excited case
in Figs.~\ref{fig:ext1} and \ref{fig:ext2}, respectively. 
\begin{figure}
\centering
 \includegraphics[width=0.79\columnwidth]{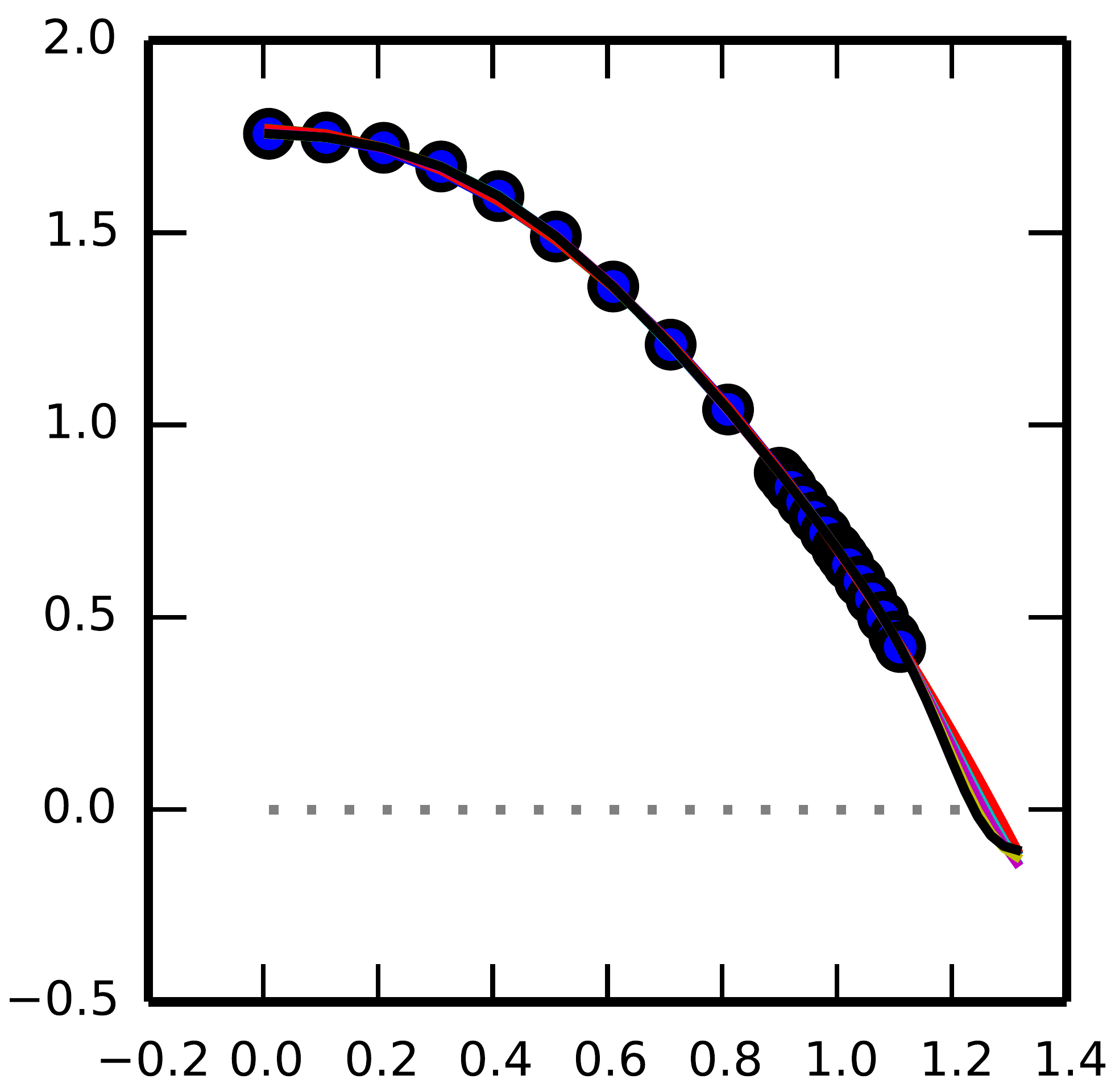}
\caption{(Color online) Same as Fig.~\ref{fig:ext1} for a scalar radial excitation.}\label{fig:ext2}      
\end{figure}
To understand the curves shown in these figures, consider the homogeneous BSE as a $P^2$-dependent eigenvalue equation
of the form
\begin{equation}\label{eq:bselambda}
\lambda(P^2)\Gamma(P^2) = \int K\; S(P^2)\; \Gamma(P^2) \;S(P^2)\;,
\end{equation}
where the original BSE is recovered for the eigenvalue $\lambda(P^2)=1$ (for more details, see \cite{Blank:2010bp}). In this
fashion, information about ground- and excited-state solutions can be extracted also off-shell and then extrapolated
to the on-shell point. We use, as provided in \cite{Blank:2011ha}, the form
\begin{equation}\label{eq:lambdapole}
\tilde{\lambda}(P^2):=\frac{\lambda(P^2)}{1 - \lambda(P^2)} = \frac{r}{P^2+M^2}+\sum_{i=1}^N \left( P^2\right)^i c_i
\end{equation}
to fit our results for the eigenvalues $\lambda$ obtained on a reasonable and directly accessible range of 
$P^2$ and straight-forwardly obtain the bound-state mass $M$ as well as the other fit constants $r$ and $c_i$. 
To understand the figures, it is important to note that, in order to reach $\lambda(P^2)=1$, we require that 
$\tilde{\lambda}(P^2)=0$. The fits are repeated with different numbers of correction 
terms $N$ in Eq.~(\ref{eq:lambdapole}), where we ensure that the fit results remain stable by a reasonable choice of 
the maximum value of $N$. Different values of $N$ yield the differently colored curves in Figs.~\ref{fig:ext1} and 
\ref{fig:ext2}, while our calculated points are depicted by black circles and the dotted line marks zero. 
We use the arithmetic mean as our final result and the differences to the 
largest and smallest values as the upper and lower error bars, as they are given in Figs.~\ref{fig:bottomonium} - 
\ref{fig:charmonium-cross}.


\begin{thebibliography}{114}%
\makeatletter
\providecommand \@ifxundefined [1]{%
 \@ifx{#1\undefined}
}%
\providecommand \@ifnum [1]{%
 \ifnum #1\expandafter \@firstoftwo
 \else \expandafter \@secondoftwo
 \fi
}%
\providecommand \@ifx [1]{%
 \ifx #1\expandafter \@firstoftwo
 \else \expandafter \@secondoftwo
 \fi
}%
\providecommand \natexlab [1]{#1}%
\providecommand \enquote  [1]{``#1''}%
\providecommand \bibnamefont  [1]{#1}%
\providecommand \bibfnamefont [1]{#1}%
\providecommand \citenamefont [1]{#1}%
\providecommand \href@noop [0]{\@secondoftwo}%
\providecommand \href [0]{\begingroup \@sanitize@url \@href}%
\providecommand \@href[1]{\@@startlink{#1}\@@href}%
\providecommand \@@href[1]{\endgroup#1\@@endlink}%
\providecommand \@sanitize@url [0]{\catcode `\\12\catcode `\$12\catcode
  `\&12\catcode `\#12\catcode `\^12\catcode `\_12\catcode `\%12\relax}%
\providecommand \@@startlink[1]{}%
\providecommand \@@endlink[0]{}%
\providecommand \url  [0]{\begingroup\@sanitize@url \@url }%
\providecommand \@url [1]{\endgroup\@href {#1}{\urlprefix }}%
\providecommand \urlprefix  [0]{URL }%
\providecommand \Eprint [0]{\href }%
\providecommand \doibase [0]{http://dx.doi.org/}%
\providecommand \selectlanguage [0]{\@gobble}%
\providecommand \bibinfo  [0]{\@secondoftwo}%
\providecommand \bibfield  [0]{\@secondoftwo}%
\providecommand \translation [1]{[#1]}%
\providecommand \BibitemOpen [0]{}%
\providecommand \bibitemStop [0]{}%
\providecommand \bibitemNoStop [0]{.\EOS\space}%
\providecommand \EOS [0]{\spacefactor3000\relax}%
\providecommand \BibitemShut  [1]{\csname bibitem#1\endcsname}%
\let\auto@bib@innerbib\@empty
\bibitem [{\citenamefont {Gross}\ and\ \citenamefont
  {Wilczek}(1973{\natexlab{a}})}]{Gross:1973id}%
  \BibitemOpen
  \bibfield  {author} {\bibinfo {author} {\bibfnamefont {D.~J.}\ \bibnamefont
  {Gross}}\ and\ \bibinfo {author} {\bibfnamefont {F.}~\bibnamefont
  {Wilczek}},\ }\href {\doibase 10.1103/PhysRevLett.30.1343} {\bibfield
  {journal} {\bibinfo  {journal} {Phys. Rev. Lett.}\ }\textbf {\bibinfo
  {volume} {30}},\ \bibinfo {pages} {1343} (\bibinfo {year}
  {1973}{\natexlab{a}})}\BibitemShut {NoStop}%
\bibitem [{\citenamefont {Gross}\ and\ \citenamefont
  {Wilczek}(1973{\natexlab{b}})}]{Gross:1973ju}%
  \BibitemOpen
  \bibfield  {author} {\bibinfo {author} {\bibfnamefont {D.~J.}\ \bibnamefont
  {Gross}}\ and\ \bibinfo {author} {\bibfnamefont {F.}~\bibnamefont
  {Wilczek}},\ }\href {\doibase 10.1103/PhysRevD.8.3633} {\bibfield  {journal}
  {\bibinfo  {journal} {Phys. Rev. D}\ }\textbf {\bibinfo {volume} {8}},\
  \bibinfo {pages} {3633} (\bibinfo {year} {1973}{\natexlab{b}})}\BibitemShut
  {NoStop}%
\bibitem [{\citenamefont {Politzer}(1973)}]{Politzer:1973fx}%
  \BibitemOpen
  \bibfield  {author} {\bibinfo {author} {\bibfnamefont {H.~D.}\ \bibnamefont
  {Politzer}},\ }\href {\doibase 10.1103/PhysRevLett.30.1346} {\bibfield
  {journal} {\bibinfo  {journal} {Phys. Rev. Lett.}\ }\textbf {\bibinfo
  {volume} {30}},\ \bibinfo {pages} {1346} (\bibinfo {year}
  {1973})}\BibitemShut {NoStop}%
\bibitem [{\citenamefont {Brambilla}\ \emph {et~al.}(2014)\citenamefont
  {Brambilla}, \citenamefont {Eidelman}, \citenamefont {Foka}, \citenamefont
  {Gardner}, \citenamefont {Kronfeld} \emph {et~al.}}]{Brambilla:2014jmp}%
  \BibitemOpen
  \bibfield  {author} {\bibinfo {author} {\bibfnamefont {N.}~\bibnamefont
  {Brambilla}}, \bibinfo {author} {\bibfnamefont {S.}~\bibnamefont {Eidelman}},
  \bibinfo {author} {\bibfnamefont {P.}~\bibnamefont {Foka}}, \bibinfo {author}
  {\bibfnamefont {S.}~\bibnamefont {Gardner}}, \bibinfo {author} {\bibfnamefont
  {A.}~\bibnamefont {Kronfeld}},  \emph {et~al.},\ }\href@noop {} {\ }\Eprint
  {http://arxiv.org/abs/1404.3723} {1404.3723 [hep-ph]} \BibitemShut {NoStop}%
\bibitem [{\citenamefont {Dudek}\ \emph {et~al.}(2008)\citenamefont {Dudek},
  \citenamefont {Edwards}, \citenamefont {Mathur},\ and\ \citenamefont
  {Richards}}]{Dudek:2007wv}%
  \BibitemOpen
  \bibfield  {author} {\bibinfo {author} {\bibfnamefont {J.~J.}\ \bibnamefont
  {Dudek}}, \bibinfo {author} {\bibfnamefont {R.~G.}\ \bibnamefont {Edwards}},
  \bibinfo {author} {\bibfnamefont {N.}~\bibnamefont {Mathur}}, \ and\ \bibinfo
  {author} {\bibfnamefont {D.~G.}\ \bibnamefont {Richards}},\ }\href {\doibase
  10.1103/PhysRevD.77.034501} {\bibfield  {journal} {\bibinfo  {journal} {Phys.
  Rev. D}\ }\textbf {\bibinfo {volume} {77}},\ \bibinfo {pages} {034501}
  (\bibinfo {year} {2008})}\BibitemShut {NoStop}%
\bibitem [{\citenamefont {Liu}\ \emph {et~al.}(2012)\citenamefont {Liu} \emph
  {et~al.}}]{Liu:2012ze}%
  \BibitemOpen
  \bibfield  {author} {\bibinfo {author} {\bibfnamefont {L.}~\bibnamefont
  {Liu}} \emph {et~al.} (\bibinfo {collaboration} {Hadron Spectrum
  Collaboration}),\ }\href {\doibase 10.1007/JHEP07(2012)126} {\bibfield
  {journal} {\bibinfo  {journal} {JHEP}\ }\textbf {\bibinfo {volume} {1207}},\
  \bibinfo {pages} {126} (\bibinfo {year} {2012})}\BibitemShut {NoStop}%
\bibitem [{\citenamefont {Thomas}(2013)}]{Thomas:2014dpa}%
  \BibitemOpen
  \bibfield  {author} {\bibinfo {author} {\bibfnamefont {C.}~\bibnamefont
  {Thomas}},\ }\href@noop {} {\bibfield  {journal} {\bibinfo  {journal} {PoS}\
  }\textbf {\bibinfo {volume} {LATTICE2013}},\ \bibinfo {pages} {003} (\bibinfo
  {year} {2013})}\BibitemShut {NoStop}%
\bibitem [{\citenamefont {Lucha}\ \emph {et~al.}(2014)\citenamefont {Lucha},
  \citenamefont {Melikhov},\ and\ \citenamefont {Simula}}]{Lucha:2014xla}%
  \BibitemOpen
  \bibfield  {author} {\bibinfo {author} {\bibfnamefont {W.}~\bibnamefont
  {Lucha}}, \bibinfo {author} {\bibfnamefont {D.}~\bibnamefont {Melikhov}}, \
  and\ \bibinfo {author} {\bibfnamefont {S.}~\bibnamefont {Simula}},\ }\href
  {\doibase 10.1016/j.physletb.2014.06.007} {\bibfield  {journal} {\bibinfo
  {journal} {Phys.Lett.}\ }\textbf {\bibinfo {volume} {B735}},\ \bibinfo
  {pages} {12} (\bibinfo {year} {2014})}\BibitemShut {NoStop}%
\bibitem [{\citenamefont {Wambach}\ \emph {et~al.}(2014)\citenamefont
  {Wambach}, \citenamefont {Tripolt}, \citenamefont {Strodthoff},\ and\
  \citenamefont {von Smekal}}]{Wambach:2014vta}%
  \BibitemOpen
  \bibfield  {author} {\bibinfo {author} {\bibfnamefont {J.}~\bibnamefont
  {Wambach}}, \bibinfo {author} {\bibfnamefont {R.-A.}\ \bibnamefont
  {Tripolt}}, \bibinfo {author} {\bibfnamefont {N.}~\bibnamefont {Strodthoff}},
  \ and\ \bibinfo {author} {\bibfnamefont {L.}~\bibnamefont {von Smekal}},\
  }\href@noop {} {\ }\Eprint {http://arxiv.org/abs/1404.7312} {1404.7312
  [hep-ph]} \BibitemShut {NoStop}%
\bibitem [{\citenamefont {Pawlowski}(2007)}]{Pawlowski:2005xe}%
  \BibitemOpen
  \bibfield  {author} {\bibinfo {author} {\bibfnamefont {J.~M.}\ \bibnamefont
  {Pawlowski}},\ }\href {\doibase 10.1016/j.aop.2007.01.007} {\bibfield
  {journal} {\bibinfo  {journal} {Annals Phys.}\ }\textbf {\bibinfo {volume}
  {322}},\ \bibinfo {pages} {2831} (\bibinfo {year} {2007})}\BibitemShut
  {NoStop}%
\bibitem [{\citenamefont {Brodsky}\ \emph {et~al.}(2014)\citenamefont
  {Brodsky}, \citenamefont {de~Teramond}, \citenamefont {Dosch},\ and\
  \citenamefont {Erlich}}]{Brodsky:2014yha}%
  \BibitemOpen
  \bibfield  {author} {\bibinfo {author} {\bibfnamefont {S.~J.}\ \bibnamefont
  {Brodsky}}, \bibinfo {author} {\bibfnamefont {G.~F.}\ \bibnamefont
  {de~Teramond}}, \bibinfo {author} {\bibfnamefont {H.~G.}\ \bibnamefont
  {Dosch}}, \ and\ \bibinfo {author} {\bibfnamefont {J.}~\bibnamefont
  {Erlich}},\ }\href@noop {} {\ }\Eprint {http://arxiv.org/abs/1407.8131}
  {1407.8131 [hep-ph]} \BibitemShut {NoStop}%
\bibitem [{\citenamefont {Roberts}(2012)}]{Roberts:2012sv}%
  \BibitemOpen
  \bibfield  {author} {\bibinfo {author} {\bibfnamefont {C.~D.}\ \bibnamefont
  {Roberts}},\ }\href@noop {} {\ }\Eprint {http://arxiv.org/abs/1203.5341}
  {1203.5341 [nucl-th]} \BibitemShut {NoStop}%
\bibitem [{\citenamefont {Bashir}\ \emph {et~al.}(2012)\citenamefont {Bashir},
  \citenamefont {Chang}, \citenamefont {Cloet}, \citenamefont {El-Bennich},
  \citenamefont {Liu} \emph {et~al.}}]{Bashir:2012fs}%
  \BibitemOpen
  \bibfield  {author} {\bibinfo {author} {\bibfnamefont {A.}~\bibnamefont
  {Bashir}}, \bibinfo {author} {\bibfnamefont {L.}~\bibnamefont {Chang}},
  \bibinfo {author} {\bibfnamefont {I.~C.}\ \bibnamefont {Cloet}}, \bibinfo
  {author} {\bibfnamefont {B.}~\bibnamefont {El-Bennich}}, \bibinfo {author}
  {\bibfnamefont {Y.-X.}\ \bibnamefont {Liu}},  \emph {et~al.},\ }\href
  {\doibase 10.1088/0253-6102/58/1/16} {\bibfield  {journal} {\bibinfo
  {journal} {Commun.Theor.Phys.}\ }\textbf {\bibinfo {volume} {58}},\ \bibinfo
  {pages} {79} (\bibinfo {year} {2012})}\BibitemShut {NoStop}%
\bibitem [{\citenamefont {Popovici}(2013)}]{Popovici:2013fya}%
  \BibitemOpen
  \bibfield  {author} {\bibinfo {author} {\bibfnamefont {C.}~\bibnamefont
  {Popovici}},\ }\href {\doibase 10.1142/S0217732313300061} {\bibfield
  {journal} {\bibinfo  {journal} {Mod.Phys.Lett.}\ }\textbf {\bibinfo {volume}
  {A28}},\ \bibinfo {pages} {1330006} (\bibinfo {year} {2013})}\BibitemShut
  {NoStop}%
\bibitem [{\citenamefont {Popovici}\ \emph {et~al.}(2010)\citenamefont
  {Popovici}, \citenamefont {Watson},\ and\ \citenamefont
  {Reinhardt}}]{Popovici:2010mb}%
  \BibitemOpen
  \bibfield  {author} {\bibinfo {author} {\bibfnamefont {C.}~\bibnamefont
  {Popovici}}, \bibinfo {author} {\bibfnamefont {P.}~\bibnamefont {Watson}}, \
  and\ \bibinfo {author} {\bibfnamefont {H.}~\bibnamefont {Reinhardt}},\ }\href
  {\doibase 10.1103/PhysRevD.81.105011} {\bibfield  {journal} {\bibinfo
  {journal} {Phys. Rev. D}\ }\textbf {\bibinfo {volume} {81}},\ \bibinfo
  {pages} {105011} (\bibinfo {year} {2010})}\BibitemShut {NoStop}%
\bibitem [{\citenamefont {Fischer}\ \emph {et~al.}(2008)\citenamefont
  {Fischer}, \citenamefont {Maas},\ and\ \citenamefont
  {Pawlowski}}]{Fischer:2008uz}%
  \BibitemOpen
  \bibfield  {author} {\bibinfo {author} {\bibfnamefont {C.~S.}\ \bibnamefont
  {Fischer}}, \bibinfo {author} {\bibfnamefont {A.}~\bibnamefont {Maas}}, \
  and\ \bibinfo {author} {\bibfnamefont {J.~M.}\ \bibnamefont {Pawlowski}},\
  }\href {\doibase 10.1016/j.aop.2009.07.009} {\bibfield  {journal} {\bibinfo
  {journal} {Annals Phys.}\ }\textbf {\bibinfo {volume} {324}},\ \bibinfo
  {pages} {2408} (\bibinfo {year} {2008})}\BibitemShut {NoStop}%
\bibitem [{\citenamefont {Bloch}\ \emph {et~al.}(2003)\citenamefont {Bloch},
  \citenamefont {Krassnigg},\ and\ \citenamefont {Roberts}}]{Bloch:2003vn}%
  \BibitemOpen
  \bibfield  {author} {\bibinfo {author} {\bibfnamefont {J.~C.~R.}\
  \bibnamefont {Bloch}}, \bibinfo {author} {\bibfnamefont {A.}~\bibnamefont
  {Krassnigg}}, \ and\ \bibinfo {author} {\bibfnamefont {C.~D.}\ \bibnamefont
  {Roberts}},\ }\href {\doibase 10.1007/s00601-003-0018-y} {\bibfield
  {journal} {\bibinfo  {journal} {Few-Body Syst.}\ }\textbf {\bibinfo {volume}
  {33}},\ \bibinfo {pages} {219} (\bibinfo {year} {2003})}\BibitemShut
  {NoStop}%
\bibitem [{\citenamefont {Eichmann}\ \emph
  {et~al.}(2008{\natexlab{a}})\citenamefont {Eichmann}, \citenamefont
  {Krassnigg}, \citenamefont {Schwinzerl},\ and\ \citenamefont
  {Alkofer}}]{Eichmann:2007nn}%
  \BibitemOpen
  \bibfield  {author} {\bibinfo {author} {\bibfnamefont {G.}~\bibnamefont
  {Eichmann}}, \bibinfo {author} {\bibfnamefont {A.}~\bibnamefont {Krassnigg}},
  \bibinfo {author} {\bibfnamefont {M.}~\bibnamefont {Schwinzerl}}, \ and\
  \bibinfo {author} {\bibfnamefont {R.}~\bibnamefont {Alkofer}},\ }\href
  {\doibase 10.1016/j.aop.2008.02.007} {\bibfield  {journal} {\bibinfo
  {journal} {Annals Phys.}\ }\textbf {\bibinfo {volume} {323}},\ \bibinfo
  {pages} {2505} (\bibinfo {year} {2008}{\natexlab{a}})}\BibitemShut {NoStop}%
\bibitem [{\citenamefont {Eichmann}\ \emph {et~al.}(2010)\citenamefont
  {Eichmann}, \citenamefont {Alkofer}, \citenamefont {Krassnigg},\ and\
  \citenamefont {Nicmorus}}]{Eichmann:2009qa}%
  \BibitemOpen
  \bibfield  {author} {\bibinfo {author} {\bibfnamefont {G.}~\bibnamefont
  {Eichmann}}, \bibinfo {author} {\bibfnamefont {R.}~\bibnamefont {Alkofer}},
  \bibinfo {author} {\bibfnamefont {A.}~\bibnamefont {Krassnigg}}, \ and\
  \bibinfo {author} {\bibfnamefont {D.}~\bibnamefont {Nicmorus}},\ }\href
  {\doibase 10.1103/PhysRevLett.104.201601} {\bibfield  {journal} {\bibinfo
  {journal} {Phys. Rev. Lett.}\ }\textbf {\bibinfo {volume} {104}},\ \bibinfo
  {pages} {201601} (\bibinfo {year} {2010})}\BibitemShut {NoStop}%
\bibitem [{\citenamefont {Sanchis-Alepuz}\ \emph {et~al.}(2011)\citenamefont
  {Sanchis-Alepuz}, \citenamefont {Eichmann}, \citenamefont {Villalba-Chavez},\
  and\ \citenamefont {Alkofer}}]{SanchisAlepuz:2011jn}%
  \BibitemOpen
  \bibfield  {author} {\bibinfo {author} {\bibfnamefont {H.}~\bibnamefont
  {Sanchis-Alepuz}}, \bibinfo {author} {\bibfnamefont {G.}~\bibnamefont
  {Eichmann}}, \bibinfo {author} {\bibfnamefont {S.}~\bibnamefont
  {Villalba-Chavez}}, \ and\ \bibinfo {author} {\bibfnamefont {R.}~\bibnamefont
  {Alkofer}},\ }\href {\doibase 10.1103/PhysRevD.84.096003} {\bibfield
  {journal} {\bibinfo  {journal} {Phys.Rev.}\ }\textbf {\bibinfo {volume}
  {D84}},\ \bibinfo {pages} {096003} (\bibinfo {year} {2011})}\BibitemShut
  {NoStop}%
\bibitem [{\citenamefont {Eichmann}\ and\ \citenamefont
  {Fischer}(2013)}]{Eichmann:2012mp}%
  \BibitemOpen
  \bibfield  {author} {\bibinfo {author} {\bibfnamefont {G.}~\bibnamefont
  {Eichmann}}\ and\ \bibinfo {author} {\bibfnamefont {C.~S.}\ \bibnamefont
  {Fischer}},\ }\href {\doibase 10.1103/PhysRevD.87.036006} {\bibfield
  {journal} {\bibinfo  {journal} {Phys.Rev.}\ }\textbf {\bibinfo {volume}
  {D87}},\ \bibinfo {pages} {036006} (\bibinfo {year} {2013})}\BibitemShut
  {NoStop}%
\bibitem [{\citenamefont {Segovia}\ \emph {et~al.}(2014)\citenamefont
  {Segovia}, \citenamefont {Chen}, \citenamefont {Cloet}, \citenamefont
  {Roberts}, \citenamefont {Schmidt} \emph {et~al.}}]{Segovia:2013uga}%
  \BibitemOpen
  \bibfield  {author} {\bibinfo {author} {\bibfnamefont {J.}~\bibnamefont
  {Segovia}}, \bibinfo {author} {\bibfnamefont {C.}~\bibnamefont {Chen}},
  \bibinfo {author} {\bibfnamefont {I.~C.}\ \bibnamefont {Cloet}}, \bibinfo
  {author} {\bibfnamefont {C.~D.}\ \bibnamefont {Roberts}}, \bibinfo {author}
  {\bibfnamefont {S.~M.}\ \bibnamefont {Schmidt}},  \emph {et~al.},\ }\href
  {\doibase 10.1007/s00601-013-0734-x} {\bibfield  {journal} {\bibinfo
  {journal} {Few Body Syst.}\ }\textbf {\bibinfo {volume} {55}},\ \bibinfo
  {pages} {1} (\bibinfo {year} {2014})}\BibitemShut {NoStop}%
\bibitem [{\citenamefont {Sanchis-Alepuz}\ \emph {et~al.}(2013)\citenamefont
  {Sanchis-Alepuz}, \citenamefont {Williams},\ and\ \citenamefont
  {Alkofer}}]{SanchisAlepuz:2013iia}%
  \BibitemOpen
  \bibfield  {author} {\bibinfo {author} {\bibfnamefont {H.}~\bibnamefont
  {Sanchis-Alepuz}}, \bibinfo {author} {\bibfnamefont {R.}~\bibnamefont
  {Williams}}, \ and\ \bibinfo {author} {\bibfnamefont {R.}~\bibnamefont
  {Alkofer}},\ }\href {\doibase 10.1103/PhysRevD.87.096015} {\bibfield
  {journal} {\bibinfo  {journal} {Phys.Rev.}\ }\textbf {\bibinfo {volume}
  {D87}},\ \bibinfo {pages} {096015} (\bibinfo {year} {2013})}\BibitemShut
  {NoStop}%
\bibitem [{\citenamefont {Sanchis-Alepuz}\ and\ \citenamefont
  {Fischer}(2014)}]{Sanchis-Alepuz:2014sca}%
  \BibitemOpen
  \bibfield  {author} {\bibinfo {author} {\bibfnamefont {H.}~\bibnamefont
  {Sanchis-Alepuz}}\ and\ \bibinfo {author} {\bibfnamefont {C.~S.}\
  \bibnamefont {Fischer}},\ }\href@noop {} {\ }\Eprint
  {http://arxiv.org/abs/1408.5577} {1408.5577 [hep-ph]} \BibitemShut {NoStop}%
\bibitem [{\citenamefont {Munczek}(1995)}]{Munczek:1994zz}%
  \BibitemOpen
  \bibfield  {author} {\bibinfo {author} {\bibfnamefont {H.~J.}\ \bibnamefont
  {Munczek}},\ }\href {\doibase 10.1103/PhysRevD.52.4736} {\bibfield  {journal}
  {\bibinfo  {journal} {Phys. Rev. D}\ }\textbf {\bibinfo {volume} {52}},\
  \bibinfo {pages} {4736} (\bibinfo {year} {1995})}\BibitemShut {NoStop}%
\bibitem [{\citenamefont {Bender}\ \emph {et~al.}(1996)\citenamefont {Bender},
  \citenamefont {Roberts},\ and\ \citenamefont {Von~Smekal}}]{Bender:1996bb}%
  \BibitemOpen
  \bibfield  {author} {\bibinfo {author} {\bibfnamefont {A.}~\bibnamefont
  {Bender}}, \bibinfo {author} {\bibfnamefont {C.~D.}\ \bibnamefont {Roberts}},
  \ and\ \bibinfo {author} {\bibfnamefont {L.}~\bibnamefont {Von~Smekal}},\
  }\href {\doibase 10.1016/0370-2693(96)00372-3} {\bibfield  {journal}
  {\bibinfo  {journal} {Phys. Lett. B}\ }\textbf {\bibinfo {volume} {380}},\
  \bibinfo {pages} {7} (\bibinfo {year} {1996})}\BibitemShut {NoStop}%
\bibitem [{\citenamefont {Bhagwat}\ \emph {et~al.}(2004)\citenamefont
  {Bhagwat}, \citenamefont {Holl}, \citenamefont {Krassnigg}, \citenamefont
  {Roberts},\ and\ \citenamefont {Tandy}}]{Bhagwat:2004hn}%
  \BibitemOpen
  \bibfield  {author} {\bibinfo {author} {\bibfnamefont {M.~S.}\ \bibnamefont
  {Bhagwat}}, \bibinfo {author} {\bibfnamefont {A.}~\bibnamefont {Holl}},
  \bibinfo {author} {\bibfnamefont {A.}~\bibnamefont {Krassnigg}}, \bibinfo
  {author} {\bibfnamefont {C.~D.}\ \bibnamefont {Roberts}}, \ and\ \bibinfo
  {author} {\bibfnamefont {P.~C.}\ \bibnamefont {Tandy}},\ }\href {\doibase
  10.1103/PhysRevC.70.035205} {\bibfield  {journal} {\bibinfo  {journal} {Phys.
  Rev. C}\ }\textbf {\bibinfo {volume} {70}},\ \bibinfo {pages} {035205}
  (\bibinfo {year} {2004})}\BibitemShut {NoStop}%
\bibitem [{\citenamefont {Maris}\ \emph {et~al.}(1998)\citenamefont {Maris},
  \citenamefont {Roberts},\ and\ \citenamefont {Tandy}}]{Maris:1997hd}%
  \BibitemOpen
  \bibfield  {author} {\bibinfo {author} {\bibfnamefont {P.}~\bibnamefont
  {Maris}}, \bibinfo {author} {\bibfnamefont {C.~D.}\ \bibnamefont {Roberts}},
  \ and\ \bibinfo {author} {\bibfnamefont {P.~C.}\ \bibnamefont {Tandy}},\
  }\href {\doibase 10.1016/S0370-2693(97)01535-9} {\bibfield  {journal}
  {\bibinfo  {journal} {Phys. Lett. B}\ }\textbf {\bibinfo {volume} {420}},\
  \bibinfo {pages} {267} (\bibinfo {year} {1998})}\BibitemShut {NoStop}%
\bibitem [{\citenamefont {Holl}\ \emph {et~al.}(2004)\citenamefont {Holl},
  \citenamefont {Krassnigg},\ and\ \citenamefont {Roberts}}]{Holl:2004fr}%
  \BibitemOpen
  \bibfield  {author} {\bibinfo {author} {\bibfnamefont {A.}~\bibnamefont
  {Holl}}, \bibinfo {author} {\bibfnamefont {A.}~\bibnamefont {Krassnigg}}, \
  and\ \bibinfo {author} {\bibfnamefont {C.~D.}\ \bibnamefont {Roberts}},\
  }\href {\doibase 10.1103/PhysRevC.70.042203} {\bibfield  {journal} {\bibinfo
  {journal} {Phys. Rev. C}\ }\textbf {\bibinfo {volume} {70}},\ \bibinfo
  {pages} {042203(R)} (\bibinfo {year} {2004})}\BibitemShut {NoStop}%
\bibitem [{\citenamefont {Maris}\ and\ \citenamefont
  {Tandy}(2000{\natexlab{a}})}]{Maris:1999bh}%
  \BibitemOpen
  \bibfield  {author} {\bibinfo {author} {\bibfnamefont {P.}~\bibnamefont
  {Maris}}\ and\ \bibinfo {author} {\bibfnamefont {P.~C.}\ \bibnamefont
  {Tandy}},\ }\href {\doibase 10.1103/PhysRevC.61.045202} {\bibfield  {journal}
  {\bibinfo  {journal} {Phys. Rev. C}\ }\textbf {\bibinfo {volume} {61}},\
  \bibinfo {pages} {045202} (\bibinfo {year} {2000}{\natexlab{a}})}\BibitemShut
  {NoStop}%
\bibitem [{\citenamefont {Maris}\ and\ \citenamefont
  {Tandy}(2006)}]{Maris:2005tt}%
  \BibitemOpen
  \bibfield  {author} {\bibinfo {author} {\bibfnamefont {P.}~\bibnamefont
  {Maris}}\ and\ \bibinfo {author} {\bibfnamefont {P.~C.}\ \bibnamefont
  {Tandy}},\ }\href {\doibase 10.1016/j.nuclphysbps.2006.08.012} {\bibfield
  {journal} {\bibinfo  {journal} {Nucl. Phys. Proc. Suppl.}\ }\textbf {\bibinfo
  {volume} {161}},\ \bibinfo {pages} {136} (\bibinfo {year}
  {2006})}\BibitemShut {NoStop}%
\bibitem [{\citenamefont {Holl}\ \emph
  {et~al.}(2005{\natexlab{a}})\citenamefont {Holl}, \citenamefont {Krassnigg},
  \citenamefont {Maris}, \citenamefont {Roberts},\ and\ \citenamefont
  {Wright}}]{Holl:2005vu}%
  \BibitemOpen
  \bibfield  {author} {\bibinfo {author} {\bibfnamefont {A.}~\bibnamefont
  {Holl}}, \bibinfo {author} {\bibfnamefont {A.}~\bibnamefont {Krassnigg}},
  \bibinfo {author} {\bibfnamefont {P.}~\bibnamefont {Maris}}, \bibinfo
  {author} {\bibfnamefont {C.~D.}\ \bibnamefont {Roberts}}, \ and\ \bibinfo
  {author} {\bibfnamefont {S.~V.}\ \bibnamefont {Wright}},\ }\href {\doibase
  10.1103/PhysRevC.71.065204} {\bibfield  {journal} {\bibinfo  {journal} {Phys.
  Rev. C}\ }\textbf {\bibinfo {volume} {71}},\ \bibinfo {pages} {065204}
  (\bibinfo {year} {2005}{\natexlab{a}})}\BibitemShut {NoStop}%
\bibitem [{\citenamefont {Bhagwat}\ and\ \citenamefont
  {Maris}(2008)}]{Bhagwat:2006pu}%
  \BibitemOpen
  \bibfield  {author} {\bibinfo {author} {\bibfnamefont {M.~S.}\ \bibnamefont
  {Bhagwat}}\ and\ \bibinfo {author} {\bibfnamefont {P.}~\bibnamefont
  {Maris}},\ }\href {\doibase 10.1103/PhysRevC.77.025203} {\bibfield  {journal}
  {\bibinfo  {journal} {Phys. Rev. C}\ }\textbf {\bibinfo {volume} {77}},\
  \bibinfo {pages} {025203} (\bibinfo {year} {2008})}\BibitemShut {NoStop}%
\bibitem [{\citenamefont {Eichmann}(2011)}]{Eichmann:2011vu}%
  \BibitemOpen
  \bibfield  {author} {\bibinfo {author} {\bibfnamefont {G.}~\bibnamefont
  {Eichmann}},\ }\href@noop {} {\bibfield  {journal} {\bibinfo  {journal}
  {Phys. Rev. D}\ }\textbf {\bibinfo {volume} {84}},\ \bibinfo {pages} {014014}
  (\bibinfo {year} {2011})}\BibitemShut {NoStop}%
\bibitem [{\citenamefont {Jarecke}\ \emph {et~al.}(2003)\citenamefont
  {Jarecke}, \citenamefont {Maris},\ and\ \citenamefont
  {Tandy}}]{Jarecke:2002xd}%
  \BibitemOpen
  \bibfield  {author} {\bibinfo {author} {\bibfnamefont {D.}~\bibnamefont
  {Jarecke}}, \bibinfo {author} {\bibfnamefont {P.}~\bibnamefont {Maris}}, \
  and\ \bibinfo {author} {\bibfnamefont {P.~C.}\ \bibnamefont {Tandy}},\ }\href
  {\doibase 10.1103/PhysRevC.67.035202} {\bibfield  {journal} {\bibinfo
  {journal} {Phys. Rev. C}\ }\textbf {\bibinfo {volume} {67}},\ \bibinfo
  {pages} {035202} (\bibinfo {year} {2003})}\BibitemShut {NoStop}%
\bibitem [{\citenamefont {Mader}\ \emph {et~al.}(2011)\citenamefont {Mader},
  \citenamefont {Eichmann}, \citenamefont {Blank},\ and\ \citenamefont
  {Krassnigg}}]{Mader:2011zf}%
  \BibitemOpen
  \bibfield  {author} {\bibinfo {author} {\bibfnamefont {V.}~\bibnamefont
  {Mader}}, \bibinfo {author} {\bibfnamefont {G.}~\bibnamefont {Eichmann}},
  \bibinfo {author} {\bibfnamefont {M.}~\bibnamefont {Blank}}, \ and\ \bibinfo
  {author} {\bibfnamefont {A.}~\bibnamefont {Krassnigg}},\ }\href@noop {}
  {\bibfield  {journal} {\bibinfo  {journal} {Phys. Rev. D}\ }\textbf {\bibinfo
  {volume} {84}},\ \bibinfo {pages} {034012} (\bibinfo {year}
  {2011})}\BibitemShut {NoStop}%
\bibitem [{\citenamefont {Nguyen}(2010)}]{Nguyen:2010ph}%
  \BibitemOpen
  \bibfield  {author} {\bibinfo {author} {\bibfnamefont {T.~T.}\ \bibnamefont
  {Nguyen}},\ }\emph {\bibinfo {title} {{Aspects of non-perturbative QCD for
  meson physics}}},\ \href@noop {} {Ph.D. thesis},\ \bibinfo  {school} {Kent
  State University} (\bibinfo {year} {2010})\BibitemShut {NoStop}%
\bibitem [{\citenamefont {Tandy}(2011)}]{Tandy:2010dw}%
  \BibitemOpen
  \bibfield  {author} {\bibinfo {author} {\bibfnamefont {P.~C.}\ \bibnamefont
  {Tandy}},\ }\href {\doibase 10.1063/1.3647112} {\bibfield  {journal}
  {\bibinfo  {journal} {AIP Conf.Proc.}\ }\textbf {\bibinfo {volume} {1374}},\
  \bibinfo {pages} {139} (\bibinfo {year} {2011})}\BibitemShut {NoStop}%
\bibitem [{\citenamefont {Nguyen}\ \emph {et~al.}(2011)\citenamefont {Nguyen},
  \citenamefont {Bashir}, \citenamefont {Roberts},\ and\ \citenamefont
  {Tandy}}]{Nguyen:2011jy}%
  \BibitemOpen
  \bibfield  {author} {\bibinfo {author} {\bibfnamefont {T.}~\bibnamefont
  {Nguyen}}, \bibinfo {author} {\bibfnamefont {A.}~\bibnamefont {Bashir}},
  \bibinfo {author} {\bibfnamefont {C.~D.}\ \bibnamefont {Roberts}}, \ and\
  \bibinfo {author} {\bibfnamefont {P.~C.}\ \bibnamefont {Tandy}},\ }\href
  {\doibase 10.1103/PhysRevC.83.062201} {\bibfield  {journal} {\bibinfo
  {journal} {Phys. Rev.}\ }\textbf {\bibinfo {volume} {C83}},\ \bibinfo {pages}
  {062201} (\bibinfo {year} {2011})}\BibitemShut {NoStop}%
\bibitem [{\citenamefont {Chang}\ \emph {et~al.}(2013)\citenamefont {Chang},
  \citenamefont {Cloet}, \citenamefont {Roberts}, \citenamefont {Schmidt},\
  and\ \citenamefont {Tandy}}]{Chang:2013nia}%
  \BibitemOpen
  \bibfield  {author} {\bibinfo {author} {\bibfnamefont {L.}~\bibnamefont
  {Chang}}, \bibinfo {author} {\bibfnamefont {I.}~\bibnamefont {Cloet}},
  \bibinfo {author} {\bibfnamefont {C.}~\bibnamefont {Roberts}}, \bibinfo
  {author} {\bibfnamefont {S.}~\bibnamefont {Schmidt}}, \ and\ \bibinfo
  {author} {\bibfnamefont {P.}~\bibnamefont {Tandy}},\ }\href {\doibase
  10.1103/PhysRevLett.111.141802} {\bibfield  {journal} {\bibinfo  {journal}
  {Phys.Rev.Lett.}\ }\textbf {\bibinfo {volume} {111}},\ \bibinfo {pages}
  {141802} (\bibinfo {year} {2013})}\BibitemShut {NoStop}%
\bibitem [{\citenamefont {Krassnigg}\ and\ \citenamefont
  {Blank}(2011)}]{Krassnigg:2010mh}%
  \BibitemOpen
  \bibfield  {author} {\bibinfo {author} {\bibfnamefont {A.}~\bibnamefont
  {Krassnigg}}\ and\ \bibinfo {author} {\bibfnamefont {M.}~\bibnamefont
  {Blank}},\ }\href {\doibase 10.1103/PhysRevD.83.096006} {\bibfield  {journal}
  {\bibinfo  {journal} {Phys. Rev. D}\ }\textbf {\bibinfo {volume} {83}},\
  \bibinfo {pages} {096006} (\bibinfo {year} {2011})}\BibitemShut {NoStop}%
\bibitem [{\citenamefont {Fischer}\ \emph
  {et~al.}(2014{\natexlab{a}})\citenamefont {Fischer}, \citenamefont {Kubrak},\
  and\ \citenamefont {Williams}}]{Fischer:2014xha}%
  \BibitemOpen
  \bibfield  {author} {\bibinfo {author} {\bibfnamefont {C.~S.}\ \bibnamefont
  {Fischer}}, \bibinfo {author} {\bibfnamefont {S.}~\bibnamefont {Kubrak}}, \
  and\ \bibinfo {author} {\bibfnamefont {R.}~\bibnamefont {Williams}},\
  }\href@noop {} {\ }\Eprint {http://arxiv.org/abs/1406.4370} {1406.4370
  [hep-ph]} \BibitemShut {NoStop}%
\bibitem [{\citenamefont {Maris}\ \emph {et~al.}(2001)\citenamefont {Maris},
  \citenamefont {Roberts}, \citenamefont {Schmidt},\ and\ \citenamefont
  {Tandy}}]{Maris:2000ig}%
  \BibitemOpen
  \bibfield  {author} {\bibinfo {author} {\bibfnamefont {P.}~\bibnamefont
  {Maris}}, \bibinfo {author} {\bibfnamefont {C.~D.}\ \bibnamefont {Roberts}},
  \bibinfo {author} {\bibfnamefont {S.~M.}\ \bibnamefont {Schmidt}}, \ and\
  \bibinfo {author} {\bibfnamefont {P.~C.}\ \bibnamefont {Tandy}},\ }\href
  {\doibase 10.1103/PhysRevC.63.025202} {\bibfield  {journal} {\bibinfo
  {journal} {Phys. Rev. C}\ }\textbf {\bibinfo {volume} {63}},\ \bibinfo
  {pages} {025202} (\bibinfo {year} {2001})}\BibitemShut {NoStop}%
\bibitem [{\citenamefont {Horvatic}\ \emph {et~al.}(2007)\citenamefont
  {Horvatic}, \citenamefont {Klabucar},\ and\ \citenamefont
  {Radzhabov}}]{Horvatic:2007qs}%
  \BibitemOpen
  \bibfield  {author} {\bibinfo {author} {\bibfnamefont {D.}~\bibnamefont
  {Horvatic}}, \bibinfo {author} {\bibfnamefont {D.}~\bibnamefont {Klabucar}},
  \ and\ \bibinfo {author} {\bibfnamefont {A.~E.}\ \bibnamefont {Radzhabov}},\
  }\href {\doibase 10.1103/PhysRevD.76.096009} {\bibfield  {journal} {\bibinfo
  {journal} {Phys. Rev. D}\ }\textbf {\bibinfo {volume} {76}},\ \bibinfo
  {pages} {096009} (\bibinfo {year} {2007})}\BibitemShut {NoStop}%
\bibitem [{\citenamefont {Blank}\ and\ \citenamefont
  {Krassnigg}(2010)}]{Blank:2010bz}%
  \BibitemOpen
  \bibfield  {author} {\bibinfo {author} {\bibfnamefont {M.}~\bibnamefont
  {Blank}}\ and\ \bibinfo {author} {\bibfnamefont {A.}~\bibnamefont
  {Krassnigg}},\ }\href {\doibase 10.1103/PhysRevD.82.034006} {\bibfield
  {journal} {\bibinfo  {journal} {Phys. Rev. D}\ }\textbf {\bibinfo {volume}
  {82}},\ \bibinfo {pages} {034006} (\bibinfo {year} {2010})}\BibitemShut
  {NoStop}%
\bibitem [{\citenamefont {Sauli}(2008)}]{Sauli:2008bn}%
  \BibitemOpen
  \bibfield  {author} {\bibinfo {author} {\bibfnamefont {V.}~\bibnamefont
  {Sauli}},\ }\href@noop {} {\bibfield  {journal} {\bibinfo  {journal} {J.
  Phys. G}\ }\textbf {\bibinfo {volume} {35}} (\bibinfo {year}
  {2008})}\BibitemShut {NoStop}%
\bibitem [{\citenamefont {Carbonell}\ and\ \citenamefont
  {Karmanov}(2010)}]{Carbonell:2010zw}%
  \BibitemOpen
  \bibfield  {author} {\bibinfo {author} {\bibfnamefont {J.}~\bibnamefont
  {Carbonell}}\ and\ \bibinfo {author} {\bibfnamefont {V.}~\bibnamefont
  {Karmanov}},\ }\href {\doibase 10.1140/epja/i2010-11055-4} {\bibfield
  {journal} {\bibinfo  {journal} {Eur.Phys.J.}\ }\textbf {\bibinfo {volume}
  {A46}},\ \bibinfo {pages} {387} (\bibinfo {year} {2010})}\BibitemShut
  {NoStop}%
\bibitem [{\citenamefont {Sauli}(2012)}]{Sauli:2011aa}%
  \BibitemOpen
  \bibfield  {author} {\bibinfo {author} {\bibfnamefont {V.}~\bibnamefont
  {Sauli}},\ }\href {\doibase 10.1103/PhysRevD.86.096004} {\bibfield  {journal}
  {\bibinfo  {journal} {Phys.Rev.}\ }\textbf {\bibinfo {volume} {D86}},\
  \bibinfo {pages} {096004} (\bibinfo {year} {2012})}\BibitemShut {NoStop}%
\bibitem [{\citenamefont {Frederico}\ \emph {et~al.}(2012)\citenamefont
  {Frederico}, \citenamefont {Salme},\ and\ \citenamefont
  {Viviani}}]{Frederico:2011ws}%
  \BibitemOpen
  \bibfield  {author} {\bibinfo {author} {\bibfnamefont {T.}~\bibnamefont
  {Frederico}}, \bibinfo {author} {\bibfnamefont {G.}~\bibnamefont {Salme}}, \
  and\ \bibinfo {author} {\bibfnamefont {M.}~\bibnamefont {Viviani}},\ }\href
  {\doibase 10.1103/PhysRevD.85.036009} {\bibfield  {journal} {\bibinfo
  {journal} {Phys.Rev.}\ }\textbf {\bibinfo {volume} {D85}},\ \bibinfo {pages}
  {036009} (\bibinfo {year} {2012})}\BibitemShut {NoStop}%
\bibitem [{\citenamefont {Sauli}(2014)}]{Sauli:2012xj}%
  \BibitemOpen
  \bibfield  {author} {\bibinfo {author} {\bibfnamefont {V.}~\bibnamefont
  {Sauli}},\ }\href@noop {} {\bibfield  {journal} {\bibinfo  {journal}
  {Phys.Rev.}\ }\textbf {\bibinfo {volume} {D90}},\ \bibinfo {pages} {016005}
  (\bibinfo {year} {2014})}\BibitemShut {NoStop}%
\bibitem [{\citenamefont {Carbonell}\ and\ \citenamefont
  {Karmanov}(2013)}]{Carbonell:2013kwa}%
  \BibitemOpen
  \bibfield  {author} {\bibinfo {author} {\bibfnamefont {J.}~\bibnamefont
  {Carbonell}}\ and\ \bibinfo {author} {\bibfnamefont {V.}~\bibnamefont
  {Karmanov}},\ }\href {\doibase 10.1016/j.physletb.2013.10.028} {\bibfield
  {journal} {\bibinfo  {journal} {Phys.Lett.}\ }\textbf {\bibinfo {volume}
  {B727}},\ \bibinfo {pages} {319} (\bibinfo {year} {2013})}\BibitemShut
  {NoStop}%
\bibitem [{\citenamefont {Frederico}\ \emph {et~al.}(2013)\citenamefont
  {Frederico}, \citenamefont {Salm\'{e}},\ and\ \citenamefont
  {Viviani}}]{Frederico:2013vga}%
  \BibitemOpen
  \bibfield  {author} {\bibinfo {author} {\bibfnamefont {T.}~\bibnamefont
  {Frederico}}, \bibinfo {author} {\bibfnamefont {G.}~\bibnamefont
  {Salm\'{e}}}, \ and\ \bibinfo {author} {\bibfnamefont {M.}~\bibnamefont
  {Viviani}},\ }\href@noop {} {\ }\Eprint {http://arxiv.org/abs/1312.0521}
  {1312.0521 [hep-ph]} \BibitemShut {NoStop}%
\bibitem [{\citenamefont {Carbonell}\ and\ \citenamefont
  {Karmanov}(2014)}]{Carbonell:2014dwa}%
  \BibitemOpen
  \bibfield  {author} {\bibinfo {author} {\bibfnamefont {J.}~\bibnamefont
  {Carbonell}}\ and\ \bibinfo {author} {\bibfnamefont {V.}~\bibnamefont
  {Karmanov}},\ }\href@noop {} {\ }\Eprint {http://arxiv.org/abs/1408.3761}
  {1408.3761 [hep-ph]} \BibitemShut {NoStop}%
\bibitem [{\citenamefont {Hall}\ and\ \citenamefont
  {Lucha}(2014)}]{Hall:2014dua}%
  \BibitemOpen
  \bibfield  {author} {\bibinfo {author} {\bibfnamefont {R.~L.}\ \bibnamefont
  {Hall}}\ and\ \bibinfo {author} {\bibfnamefont {W.}~\bibnamefont {Lucha}},\
  }\href@noop {} {\ }\Eprint {http://arxiv.org/abs/1408.6330} {1408.6330
  [math-ph]} \BibitemShut {NoStop}%
\bibitem [{\citenamefont {Alkofer}\ \emph {et~al.}(2006)\citenamefont
  {Alkofer}, \citenamefont {Kloker}, \citenamefont {Krassnigg},\ and\
  \citenamefont {Wagenbrunn}}]{Alkofer:2005ug}%
  \BibitemOpen
  \bibfield  {author} {\bibinfo {author} {\bibfnamefont {R.}~\bibnamefont
  {Alkofer}}, \bibinfo {author} {\bibfnamefont {M.}~\bibnamefont {Kloker}},
  \bibinfo {author} {\bibfnamefont {A.}~\bibnamefont {Krassnigg}}, \ and\
  \bibinfo {author} {\bibfnamefont {R.~F.}\ \bibnamefont {Wagenbrunn}},\ }\href
  {\doibase 10.1103/PhysRevLett.96.022001} {\bibfield  {journal} {\bibinfo
  {journal} {Phys. Rev. Lett.}\ }\textbf {\bibinfo {volume} {96}},\ \bibinfo
  {pages} {022001} (\bibinfo {year} {2006})}\BibitemShut {NoStop}%
\bibitem [{\citenamefont {Gomez-Rocha}\ \emph {et~al.}(2010)\citenamefont
  {Gomez-Rocha}, \citenamefont {Llanes-Estrada}, \citenamefont {Schutte},\ and\
  \citenamefont {Villalba-Chavez}}]{Rocha:2009xq}%
  \BibitemOpen
  \bibfield  {author} {\bibinfo {author} {\bibfnamefont {M.}~\bibnamefont
  {Gomez-Rocha}}, \bibinfo {author} {\bibfnamefont {F.~J.}\ \bibnamefont
  {Llanes-Estrada}}, \bibinfo {author} {\bibfnamefont {D.}~\bibnamefont
  {Schutte}}, \ and\ \bibinfo {author} {\bibfnamefont {S.}~\bibnamefont
  {Villalba-Chavez}},\ }\href {\doibase 10.1140/epja/i2010-10949-3} {\bibfield
  {journal} {\bibinfo  {journal} {Eur.J.Phys.}\ }\textbf {\bibinfo {volume}
  {A44}},\ \bibinfo {pages} {411} (\bibinfo {year} {2010})}\BibitemShut
  {NoStop}%
\bibitem [{\citenamefont {Llanes-Estrada}\ \emph {et~al.}(2011)\citenamefont
  {Llanes-Estrada} \emph {et~al.}}]{LlanesEstrada:2010bs}%
  \BibitemOpen
  \bibfield  {author} {\bibinfo {author} {\bibfnamefont {F.~J.}\ \bibnamefont
  {Llanes-Estrada}} \emph {et~al.},\ }\href@noop {} {\bibfield  {journal}
  {\bibinfo  {journal} {Fizika}\ }\textbf {\bibinfo {volume} {B20}},\ \bibinfo
  {pages} {63} (\bibinfo {year} {2011})}\BibitemShut {NoStop}%
\bibitem [{\citenamefont {Cotanch}\ and\ \citenamefont
  {Llanes-Estrada}(2011)}]{Cotanch:2010bq}%
  \BibitemOpen
  \bibfield  {author} {\bibinfo {author} {\bibfnamefont {S.~R.}\ \bibnamefont
  {Cotanch}}\ and\ \bibinfo {author} {\bibfnamefont {F.~J.}\ \bibnamefont
  {Llanes-Estrada}},\ }\href@noop {} {\bibfield  {journal} {\bibinfo  {journal}
  {Fizika}\ }\textbf {\bibinfo {volume} {B20}},\ \bibinfo {pages} {1} (\bibinfo
  {year} {2011})}\BibitemShut {NoStop}%
\bibitem [{\citenamefont {Popovici}\ \emph {et~al.}(2011)\citenamefont
  {Popovici}, \citenamefont {Watson},\ and\ \citenamefont
  {Reinhardt}}]{Popovici:2011yz}%
  \BibitemOpen
  \bibfield  {author} {\bibinfo {author} {\bibfnamefont {C.}~\bibnamefont
  {Popovici}}, \bibinfo {author} {\bibfnamefont {P.}~\bibnamefont {Watson}}, \
  and\ \bibinfo {author} {\bibfnamefont {H.}~\bibnamefont {Reinhardt}},\ }\href
  {\doibase 10.1103/PhysRevD.83.125018} {\bibfield  {journal} {\bibinfo
  {journal} {Phys. Rev.}\ }\textbf {\bibinfo {volume} {D83}},\ \bibinfo {pages}
  {125018} (\bibinfo {year} {2011})}\BibitemShut {NoStop}%
\bibitem [{\citenamefont {Popovici}(2011)}]{Popovici:2011wx}%
  \BibitemOpen
  \bibfield  {author} {\bibinfo {author} {\bibfnamefont {C.}~\bibnamefont
  {Popovici}},\ }\emph {\bibinfo {title} {{Quark sector of Coulomb gauge
  Quantum Chromodynamics}}},\ \href@noop {} {Ph.D. thesis},\ \bibinfo  {school}
  {University of Tuebingen} (\bibinfo {year} {2011}),\ \Eprint
  {http://arxiv.org/abs/1106.0691} {1106.0691 [hep-ph]} \BibitemShut {NoStop}%
\bibitem [{\citenamefont {Gomez-Rocha}\ \emph {et~al.}(2014)\citenamefont
  {Gomez-Rocha}, \citenamefont {Hilger},\ and\ \citenamefont
  {Krassnigg}}]{Gomez-Rocha:2014vsa}%
  \BibitemOpen
  \bibfield  {author} {\bibinfo {author} {\bibfnamefont {M.}~\bibnamefont
  {Gomez-Rocha}}, \bibinfo {author} {\bibfnamefont {T.}~\bibnamefont {Hilger}},
  \ and\ \bibinfo {author} {\bibfnamefont {A.}~\bibnamefont {Krassnigg}},\
  }\href@noop {} {\ }\Eprint {http://arxiv.org/abs/1408.1077} {1408.1077
  [hep-ph]} \BibitemShut {NoStop}%
\bibitem [{\citenamefont {Watson}\ and\ \citenamefont
  {Cassing}(2004)}]{Watson:2004jq}%
  \BibitemOpen
  \bibfield  {author} {\bibinfo {author} {\bibfnamefont {P.}~\bibnamefont
  {Watson}}\ and\ \bibinfo {author} {\bibfnamefont {W.}~\bibnamefont
  {Cassing}},\ }\href {\doibase 10.1007/s00601-004-0063-1} {\bibfield
  {journal} {\bibinfo  {journal} {Few-Body Syst.}\ }\textbf {\bibinfo {volume}
  {35}},\ \bibinfo {pages} {99} (\bibinfo {year} {2004})}\BibitemShut {NoStop}%
\bibitem [{\citenamefont {Watson}\ \emph {et~al.}(2004)\citenamefont {Watson},
  \citenamefont {Cassing},\ and\ \citenamefont {Tandy}}]{Watson:2004kd}%
  \BibitemOpen
  \bibfield  {author} {\bibinfo {author} {\bibfnamefont {P.}~\bibnamefont
  {Watson}}, \bibinfo {author} {\bibfnamefont {W.}~\bibnamefont {Cassing}}, \
  and\ \bibinfo {author} {\bibfnamefont {P.~C.}\ \bibnamefont {Tandy}},\ }\href
  {\doibase 10.1007/s00601-004-0067-x} {\bibfield  {journal} {\bibinfo
  {journal} {Few-Body Syst.}\ }\textbf {\bibinfo {volume} {35}},\ \bibinfo
  {pages} {129} (\bibinfo {year} {2004})}\BibitemShut {NoStop}%
\bibitem [{\citenamefont {Fischer}\ \emph {et~al.}(2005)\citenamefont
  {Fischer}, \citenamefont {Watson},\ and\ \citenamefont
  {Cassing}}]{Fischer:2005en}%
  \BibitemOpen
  \bibfield  {author} {\bibinfo {author} {\bibfnamefont {C.~S.}\ \bibnamefont
  {Fischer}}, \bibinfo {author} {\bibfnamefont {P.}~\bibnamefont {Watson}}, \
  and\ \bibinfo {author} {\bibfnamefont {W.}~\bibnamefont {Cassing}},\ }\href
  {\doibase 10.1103/PhysRevD.72.094025} {\bibfield  {journal} {\bibinfo
  {journal} {Phys. Rev. D}\ }\textbf {\bibinfo {volume} {72}},\ \bibinfo
  {pages} {094025} (\bibinfo {year} {2005})}\BibitemShut {NoStop}%
\bibitem [{\citenamefont {Matevosyan}\ \emph
  {et~al.}(2007{\natexlab{a}})\citenamefont {Matevosyan}, \citenamefont
  {Thomas},\ and\ \citenamefont {Tandy}}]{Matevosyan:2006bk}%
  \BibitemOpen
  \bibfield  {author} {\bibinfo {author} {\bibfnamefont {H.~H.}\ \bibnamefont
  {Matevosyan}}, \bibinfo {author} {\bibfnamefont {A.~W.}\ \bibnamefont
  {Thomas}}, \ and\ \bibinfo {author} {\bibfnamefont {P.~C.}\ \bibnamefont
  {Tandy}},\ }\href@noop {} {\bibfield  {journal} {\bibinfo  {journal} {Phys.
  Rev. C}\ }\textbf {\bibinfo {volume} {75}},\ \bibinfo {pages} {045201}
  (\bibinfo {year} {2007}{\natexlab{a}})}\BibitemShut {NoStop}%
\bibitem [{\citenamefont {Matevosyan}\ \emph
  {et~al.}(2007{\natexlab{b}})\citenamefont {Matevosyan}, \citenamefont
  {Thomas},\ and\ \citenamefont {Tandy}}]{Matevosyan:2007cx}%
  \BibitemOpen
  \bibfield  {author} {\bibinfo {author} {\bibfnamefont {H.~H.}\ \bibnamefont
  {Matevosyan}}, \bibinfo {author} {\bibfnamefont {A.~W.}\ \bibnamefont
  {Thomas}}, \ and\ \bibinfo {author} {\bibfnamefont {P.~C.}\ \bibnamefont
  {Tandy}},\ }\href {\doibase 10.1088/0954-3899/34/10/005} {\bibfield
  {journal} {\bibinfo  {journal} {J. Phys. G}\ }\textbf {\bibinfo {volume}
  {34}},\ \bibinfo {pages} {2153} (\bibinfo {year}
  {2007}{\natexlab{b}})}\BibitemShut {NoStop}%
\bibitem [{\citenamefont {Fischer}\ and\ \citenamefont
  {Williams}(2008)}]{Fischer:2008wy}%
  \BibitemOpen
  \bibfield  {author} {\bibinfo {author} {\bibfnamefont {C.~S.}\ \bibnamefont
  {Fischer}}\ and\ \bibinfo {author} {\bibfnamefont {R.}~\bibnamefont
  {Williams}},\ }\href {\doibase 10.1103/PhysRevD.78.074006} {\bibfield
  {journal} {\bibinfo  {journal} {Phys. Rev. D}\ }\textbf {\bibinfo {volume}
  {78}},\ \bibinfo {pages} {074006} (\bibinfo {year} {2008})}\BibitemShut
  {NoStop}%
\bibitem [{\citenamefont {Fischer}\ and\ \citenamefont
  {Williams}(2009)}]{Fischer:2009jm}%
  \BibitemOpen
  \bibfield  {author} {\bibinfo {author} {\bibfnamefont {C.~S.}\ \bibnamefont
  {Fischer}}\ and\ \bibinfo {author} {\bibfnamefont {R.}~\bibnamefont
  {Williams}},\ }\href {\doibase 10.1103/PhysRevLett.103.122001} {\bibfield
  {journal} {\bibinfo  {journal} {Phys. Rev. Lett.}\ }\textbf {\bibinfo
  {volume} {103}},\ \bibinfo {pages} {122001} (\bibinfo {year}
  {2009})}\BibitemShut {NoStop}%
\bibitem [{\citenamefont {Williams}(2010)}]{Williams:2009wx}%
  \BibitemOpen
  \bibfield  {author} {\bibinfo {author} {\bibfnamefont {R.}~\bibnamefont
  {Williams}},\ }\href {\doibase 10.1051/epjconf/20100303005} {\bibfield
  {journal} {\bibinfo  {journal} {EPJ Web Conf.}\ }\textbf {\bibinfo {volume}
  {3}},\ \bibinfo {pages} {03005} (\bibinfo {year} {2010})}\BibitemShut
  {NoStop}%
\bibitem [{\citenamefont {Williams}(2014)}]{Williams:2014iea}%
  \BibitemOpen
  \bibfield  {author} {\bibinfo {author} {\bibfnamefont {R.}~\bibnamefont
  {Williams}},\ }\href@noop {} {\ }\Eprint {http://arxiv.org/abs/1404.2545}
  {1404.2545 [hep-ph]} \BibitemShut {NoStop}%
\bibitem [{\citenamefont {Sanchis-Alepuz}\ \emph {et~al.}(2014)\citenamefont
  {Sanchis-Alepuz}, \citenamefont {Fischer},\ and\ \citenamefont
  {Kubrak}}]{Sanchis-Alepuz:2014wea}%
  \BibitemOpen
  \bibfield  {author} {\bibinfo {author} {\bibfnamefont {H.}~\bibnamefont
  {Sanchis-Alepuz}}, \bibinfo {author} {\bibfnamefont {C.~S.}\ \bibnamefont
  {Fischer}}, \ and\ \bibinfo {author} {\bibfnamefont {S.}~\bibnamefont
  {Kubrak}},\ }\href@noop {} {\ }\Eprint {http://arxiv.org/abs/1401.3183}
  {1401.3183 [hep-ph]} \BibitemShut {NoStop}%
\bibitem [{\citenamefont {Chang}\ and\ \citenamefont
  {Roberts}(2009)}]{Chang:2009zb}%
  \BibitemOpen
  \bibfield  {author} {\bibinfo {author} {\bibfnamefont {L.}~\bibnamefont
  {Chang}}\ and\ \bibinfo {author} {\bibfnamefont {C.~D.}\ \bibnamefont
  {Roberts}},\ }\href {\doibase 10.1103/PhysRevLett.103.081601} {\bibfield
  {journal} {\bibinfo  {journal} {Phys. Rev. Lett.}\ }\textbf {\bibinfo
  {volume} {103}},\ \bibinfo {pages} {081601} (\bibinfo {year}
  {2009})}\BibitemShut {NoStop}%
\bibitem [{\citenamefont {Heupel}\ \emph {et~al.}(2014)\citenamefont {Heupel},
  \citenamefont {Goecke},\ and\ \citenamefont {Fischer}}]{Heupel:2014ina}%
  \BibitemOpen
  \bibfield  {author} {\bibinfo {author} {\bibfnamefont {W.}~\bibnamefont
  {Heupel}}, \bibinfo {author} {\bibfnamefont {T.}~\bibnamefont {Goecke}}, \
  and\ \bibinfo {author} {\bibfnamefont {C.~S.}\ \bibnamefont {Fischer}},\
  }\href {\doibase 10.1140/epja/i2014-14085-x} {\bibfield  {journal} {\bibinfo
  {journal} {Eur.Phys.J.}\ }\textbf {\bibinfo {volume} {A50}},\ \bibinfo
  {pages} {85} (\bibinfo {year} {2014})}\BibitemShut {NoStop}%
\bibitem [{\citenamefont {Maskawa}\ and\ \citenamefont
  {Nakajima}(1974)}]{Maskawa:1974vs}%
  \BibitemOpen
  \bibfield  {author} {\bibinfo {author} {\bibfnamefont {T.}~\bibnamefont
  {Maskawa}}\ and\ \bibinfo {author} {\bibfnamefont {H.}~\bibnamefont
  {Nakajima}},\ }\href {http://ptp.ipap.jp/link?PTP/52/1326} {\bibfield
  {journal} {\bibinfo  {journal} {Prog. Theor. Phys.}\ }\textbf {\bibinfo
  {volume} {52}},\ \bibinfo {pages} {1326} (\bibinfo {year}
  {1974})}\BibitemShut {NoStop}%
\bibitem [{\citenamefont {Aoki}\ \emph {et~al.}(1990)\citenamefont {Aoki},
  \citenamefont {Bando}, \citenamefont {Kugo}, \citenamefont {Mitchard},\ and\
  \citenamefont {Nakatani}}]{Aoki:1990eq}%
  \BibitemOpen
  \bibfield  {author} {\bibinfo {author} {\bibfnamefont {K.-I.}\ \bibnamefont
  {Aoki}}, \bibinfo {author} {\bibfnamefont {M.}~\bibnamefont {Bando}},
  \bibinfo {author} {\bibfnamefont {T.}~\bibnamefont {Kugo}}, \bibinfo {author}
  {\bibfnamefont {M.~G.}\ \bibnamefont {Mitchard}}, \ and\ \bibinfo {author}
  {\bibfnamefont {H.}~\bibnamefont {Nakatani}},\ }\href {\doibase
  10.1143/PTP.84.683} {\bibfield  {journal} {\bibinfo  {journal} {Prog. Theor.
  Phys.}\ }\textbf {\bibinfo {volume} {84}},\ \bibinfo {pages} {683} (\bibinfo
  {year} {1990})}\BibitemShut {NoStop}%
\bibitem [{\citenamefont {Kugo}\ and\ \citenamefont
  {Mitchard}(1992{\natexlab{a}})}]{Kugo:1992pr}%
  \BibitemOpen
  \bibfield  {author} {\bibinfo {author} {\bibfnamefont {T.}~\bibnamefont
  {Kugo}}\ and\ \bibinfo {author} {\bibfnamefont {M.~G.}\ \bibnamefont
  {Mitchard}},\ }\href {\doibase 10.1016/0370-2693(92)90496-Q} {\bibfield
  {journal} {\bibinfo  {journal} {Phys. Lett. B}\ }\textbf {\bibinfo {volume}
  {282}},\ \bibinfo {pages} {162} (\bibinfo {year}
  {1992}{\natexlab{a}})}\BibitemShut {NoStop}%
\bibitem [{\citenamefont {Bando}\ \emph {et~al.}(1994)\citenamefont {Bando},
  \citenamefont {Harada},\ and\ \citenamefont {Kugo}}]{Bando:1993qy}%
  \BibitemOpen
  \bibfield  {author} {\bibinfo {author} {\bibfnamefont {M.}~\bibnamefont
  {Bando}}, \bibinfo {author} {\bibfnamefont {M.}~\bibnamefont {Harada}}, \
  and\ \bibinfo {author} {\bibfnamefont {T.}~\bibnamefont {Kugo}},\ }\href
  {\doibase 10.1143/PTP.91.927} {\bibfield  {journal} {\bibinfo  {journal}
  {Prog. Theor. Phys.}\ }\textbf {\bibinfo {volume} {91}},\ \bibinfo {pages}
  {927} (\bibinfo {year} {1994})}\BibitemShut {NoStop}%
\bibitem [{\citenamefont {Maris}\ and\ \citenamefont
  {Tandy}(2000{\natexlab{b}})}]{Maris:2000sk}%
  \BibitemOpen
  \bibfield  {author} {\bibinfo {author} {\bibfnamefont {P.}~\bibnamefont
  {Maris}}\ and\ \bibinfo {author} {\bibfnamefont {P.~C.}\ \bibnamefont
  {Tandy}},\ }\href {\doibase 10.1103/PhysRevC.62.055204} {\bibfield  {journal}
  {\bibinfo  {journal} {Phys. Rev. C}\ }\textbf {\bibinfo {volume} {62}},\
  \bibinfo {pages} {055204} (\bibinfo {year} {2000}{\natexlab{b}})}\BibitemShut
  {NoStop}%
\bibitem [{\citenamefont {Krassnigg}(2008)}]{Krassnigg:2008gd}%
  \BibitemOpen
  \bibfield  {author} {\bibinfo {author} {\bibfnamefont {A.}~\bibnamefont
  {Krassnigg}},\ }\href@noop {} {\bibfield  {journal} {\bibinfo  {journal}
  {PoS}\ }\textbf {\bibinfo {volume} {Confinement8}},\ \bibinfo {pages} {075}
  (\bibinfo {year} {2008})}\BibitemShut {NoStop}%
\bibitem [{\citenamefont {Blank}\ and\ \citenamefont
  {Krassnigg}(2011{\natexlab{a}})}]{Blank:2010bp}%
  \BibitemOpen
  \bibfield  {author} {\bibinfo {author} {\bibfnamefont {M.}~\bibnamefont
  {Blank}}\ and\ \bibinfo {author} {\bibfnamefont {A.}~\bibnamefont
  {Krassnigg}},\ }\href {\doibase 10.1016/j.cpc.2011.03.003} {\bibfield
  {journal} {\bibinfo  {journal} {Comput. Phys. Commun.}\ }\textbf {\bibinfo
  {volume} {182}},\ \bibinfo {pages} {1391} (\bibinfo {year}
  {2011}{\natexlab{a}})}\BibitemShut {NoStop}%
\bibitem [{\citenamefont {Blank}(2011)}]{Blank:2011qk}%
  \BibitemOpen
  \bibfield  {author} {\bibinfo {author} {\bibfnamefont {M.}~\bibnamefont
  {Blank}},\ }\emph {\bibinfo {title} {{Properties of quarks and mesons in the
  Dyson-Schwinger/Bethe-Salpeter approach}}},\ \href@noop {} {Ph.D. thesis},\
  \bibinfo  {school} {University of Graz} (\bibinfo {year} {2011}),\ \Eprint
  {http://arxiv.org/abs/1106.4843} {1106.4843 [hep-ph]} \BibitemShut {NoStop}%
\bibitem [{\citenamefont {Bhagwat}\ \emph {et~al.}(2007)\citenamefont
  {Bhagwat}, \citenamefont {Hoell}, \citenamefont {Krassnigg}, \citenamefont
  {Roberts},\ and\ \citenamefont {Wright}}]{Bhagwat:2007rj}%
  \BibitemOpen
  \bibfield  {author} {\bibinfo {author} {\bibfnamefont {M.~S.}\ \bibnamefont
  {Bhagwat}}, \bibinfo {author} {\bibfnamefont {A.}~\bibnamefont {Hoell}},
  \bibinfo {author} {\bibfnamefont {A.}~\bibnamefont {Krassnigg}}, \bibinfo
  {author} {\bibfnamefont {C.~D.}\ \bibnamefont {Roberts}}, \ and\ \bibinfo
  {author} {\bibfnamefont {S.~V.}\ \bibnamefont {Wright}},\ }\href {\doibase
  10.1007/s00601-007-0174-6} {\bibfield  {journal} {\bibinfo  {journal}
  {Few-Body Syst.}\ }\textbf {\bibinfo {volume} {40}},\ \bibinfo {pages} {209}
  (\bibinfo {year} {2007})}\BibitemShut {NoStop}%
\bibitem [{\citenamefont {Blank}\ and\ \citenamefont
  {Krassnigg}(2011{\natexlab{b}})}]{Blank:2010sn}%
  \BibitemOpen
  \bibfield  {author} {\bibinfo {author} {\bibfnamefont {M.}~\bibnamefont
  {Blank}}\ and\ \bibinfo {author} {\bibfnamefont {A.}~\bibnamefont
  {Krassnigg}},\ }\href {\doibase 10.1063/1.3575026} {\bibfield  {journal}
  {\bibinfo  {journal} {AIP Conf. Proc.}\ }\textbf {\bibinfo {volume} {1343}},\
  \bibinfo {pages} {349} (\bibinfo {year} {2011}{\natexlab{b}})}\BibitemShut
  {NoStop}%
\bibitem [{\citenamefont {Maris}\ and\ \citenamefont
  {Roberts}(1997)}]{Maris:1997tm}%
  \BibitemOpen
  \bibfield  {author} {\bibinfo {author} {\bibfnamefont {P.}~\bibnamefont
  {Maris}}\ and\ \bibinfo {author} {\bibfnamefont {C.~D.}\ \bibnamefont
  {Roberts}},\ }\href {\doibase 10.1103/PhysRevC.56.3369} {\bibfield  {journal}
  {\bibinfo  {journal} {Phys. Rev. C}\ }\textbf {\bibinfo {volume} {56}},\
  \bibinfo {pages} {3369} (\bibinfo {year} {1997})}\BibitemShut {NoStop}%
\bibitem [{\citenamefont {Munczek}\ and\ \citenamefont
  {Nemirovsky}(1983)}]{Munczek:1983dx}%
  \BibitemOpen
  \bibfield  {author} {\bibinfo {author} {\bibfnamefont {H.~J.}\ \bibnamefont
  {Munczek}}\ and\ \bibinfo {author} {\bibfnamefont {A.~M.}\ \bibnamefont
  {Nemirovsky}},\ }\href {\doibase 10.1103/PhysRevD.28.181} {\bibfield
  {journal} {\bibinfo  {journal} {Phys. Rev. D}\ }\textbf {\bibinfo {volume}
  {28}},\ \bibinfo {pages} {181} (\bibinfo {year} {1983})}\BibitemShut
  {NoStop}%
\bibitem [{\citenamefont {Munczek}\ and\ \citenamefont
  {Jain}(1992)}]{Munczek:1991jb}%
  \BibitemOpen
  \bibfield  {author} {\bibinfo {author} {\bibfnamefont {H.~J.}\ \bibnamefont
  {Munczek}}\ and\ \bibinfo {author} {\bibfnamefont {P.}~\bibnamefont {Jain}},\
  }\href {\doibase 10.1103/PhysRevD.46.438} {\bibfield  {journal} {\bibinfo
  {journal} {Phys. Rev. D}\ }\textbf {\bibinfo {volume} {46}},\ \bibinfo
  {pages} {438} (\bibinfo {year} {1992})}\BibitemShut {NoStop}%
\bibitem [{\citenamefont {Jain}\ and\ \citenamefont
  {Munczek}(1993)}]{Jain:1993qh}%
  \BibitemOpen
  \bibfield  {author} {\bibinfo {author} {\bibfnamefont {P.}~\bibnamefont
  {Jain}}\ and\ \bibinfo {author} {\bibfnamefont {H.~J.}\ \bibnamefont
  {Munczek}},\ }\href {\doibase 10.1103/PhysRevD.48.5403} {\bibfield  {journal}
  {\bibinfo  {journal} {Phys. Rev. D}\ }\textbf {\bibinfo {volume} {48}},\
  \bibinfo {pages} {5403} (\bibinfo {year} {1993})}\BibitemShut {NoStop}%
\bibitem [{\citenamefont {Richardson}(1979)}]{Richardson:1978bt}%
  \BibitemOpen
  \bibfield  {author} {\bibinfo {author} {\bibfnamefont {J.~L.}\ \bibnamefont
  {Richardson}},\ }\href {\doibase 10.1016/0370-2693(79)90753-6} {\bibfield
  {journal} {\bibinfo  {journal} {Phys.Lett.}\ }\textbf {\bibinfo {volume}
  {B82}},\ \bibinfo {pages} {272} (\bibinfo {year} {1979})}\BibitemShut
  {NoStop}%
\bibitem [{\citenamefont {Higashijima}(1984)}]{Higashijima:1983gx}%
  \BibitemOpen
  \bibfield  {author} {\bibinfo {author} {\bibfnamefont {K.}~\bibnamefont
  {Higashijima}},\ }\href {\doibase 10.1103/PhysRevD.29.1228} {\bibfield
  {journal} {\bibinfo  {journal} {Phys.Rev. D}\ }\textbf {\bibinfo {volume}
  {29}},\ \bibinfo {pages} {1228} (\bibinfo {year} {1984})}\BibitemShut
  {NoStop}%
\bibitem [{\citenamefont {Higashijima}(1991)}]{Higashijima:1991de}%
  \BibitemOpen
  \bibfield  {author} {\bibinfo {author} {\bibfnamefont {K.}~\bibnamefont
  {Higashijima}},\ }\href {\doibase 10.1143/PTPS.104.1} {\bibfield  {journal}
  {\bibinfo  {journal} {Prog.Theor.Phys.Suppl.}\ }\textbf {\bibinfo {volume}
  {104}},\ \bibinfo {pages} {1} (\bibinfo {year} {1991})}\BibitemShut {NoStop}%
\bibitem [{\citenamefont {Kugo}\ and\ \citenamefont
  {Mitchard}(1992{\natexlab{b}})}]{Kugo:1992zg}%
  \BibitemOpen
  \bibfield  {author} {\bibinfo {author} {\bibfnamefont {T.}~\bibnamefont
  {Kugo}}\ and\ \bibinfo {author} {\bibfnamefont {M.~G.}\ \bibnamefont
  {Mitchard}},\ }\href {\doibase 10.1016/0370-2693(92)91787-A} {\bibfield
  {journal} {\bibinfo  {journal} {Phys. Lett. B}\ }\textbf {\bibinfo {volume}
  {286}},\ \bibinfo {pages} {355} (\bibinfo {year}
  {1992}{\natexlab{b}})}\BibitemShut {NoStop}%
\bibitem [{\citenamefont {Yamanaka}\ \emph {et~al.}(2013)\citenamefont
  {Yamanaka}, \citenamefont {Doi}, \citenamefont {Imai},\ and\ \citenamefont
  {Suganuma}}]{Yamanaka:2013zoa}%
  \BibitemOpen
  \bibfield  {author} {\bibinfo {author} {\bibfnamefont {N.}~\bibnamefont
  {Yamanaka}}, \bibinfo {author} {\bibfnamefont {T.~M.}\ \bibnamefont {Doi}},
  \bibinfo {author} {\bibfnamefont {S.}~\bibnamefont {Imai}}, \ and\ \bibinfo
  {author} {\bibfnamefont {H.}~\bibnamefont {Suganuma}},\ }\href {\doibase
  10.1103/PhysRevD.88.074036} {\bibfield  {journal} {\bibinfo  {journal}
  {Phys.Rev. D}\ }\textbf {\bibinfo {volume} {88}},\ \bibinfo {pages} {074036}
  (\bibinfo {year} {2013})}\BibitemShut {NoStop}%
\bibitem [{\citenamefont {Maris}\ and\ \citenamefont
  {Tandy}(1999)}]{Maris:1999nt}%
  \BibitemOpen
  \bibfield  {author} {\bibinfo {author} {\bibfnamefont {P.}~\bibnamefont
  {Maris}}\ and\ \bibinfo {author} {\bibfnamefont {P.~C.}\ \bibnamefont
  {Tandy}},\ }\href {\doibase 10.1103/PhysRevC.60.055214} {\bibfield  {journal}
  {\bibinfo  {journal} {Phys. Rev. C}\ }\textbf {\bibinfo {volume} {60}},\
  \bibinfo {pages} {055214} (\bibinfo {year} {1999})}\BibitemShut {NoStop}%
\bibitem [{\citenamefont {Qin}\ \emph {et~al.}(2012)\citenamefont {Qin},
  \citenamefont {Chang}, \citenamefont {Liu}, \citenamefont {Roberts},\ and\
  \citenamefont {Wilson}}]{Qin:2011xq}%
  \BibitemOpen
  \bibfield  {author} {\bibinfo {author} {\bibfnamefont {S.-X.}\ \bibnamefont
  {Qin}}, \bibinfo {author} {\bibfnamefont {L.}~\bibnamefont {Chang}}, \bibinfo
  {author} {\bibfnamefont {Y.-X.}\ \bibnamefont {Liu}}, \bibinfo {author}
  {\bibfnamefont {C.~D.}\ \bibnamefont {Roberts}}, \ and\ \bibinfo {author}
  {\bibfnamefont {D.~J.}\ \bibnamefont {Wilson}},\ }\href {\doibase
  10.1103/PhysRevC.85.035202} {\bibfield  {journal} {\bibinfo  {journal}
  {Phys.Rev.}\ }\textbf {\bibinfo {volume} {C85}},\ \bibinfo {pages} {035202}
  (\bibinfo {year} {2012})}\BibitemShut {NoStop}%
\bibitem [{\citenamefont {Fischer}\ \emph
  {et~al.}(2014{\natexlab{b}})\citenamefont {Fischer}, \citenamefont {Kubrak},\
  and\ \citenamefont {Williams}}]{Fischer:2014cfa}%
  \BibitemOpen
  \bibfield  {author} {\bibinfo {author} {\bibfnamefont {C.~S.}\ \bibnamefont
  {Fischer}}, \bibinfo {author} {\bibfnamefont {S.}~\bibnamefont {Kubrak}}, \
  and\ \bibinfo {author} {\bibfnamefont {R.}~\bibnamefont {Williams}},\
  }\href@noop {} {\ }\Eprint {http://arxiv.org/abs/1409.5076} {1409.5076
  [hep-ph]} \BibitemShut {NoStop}%
\bibitem [{\citenamefont {Frank}\ and\ \citenamefont
  {Roberts}(1996)}]{Frank:1995uk}%
  \BibitemOpen
  \bibfield  {author} {\bibinfo {author} {\bibfnamefont {M.~R.}\ \bibnamefont
  {Frank}}\ and\ \bibinfo {author} {\bibfnamefont {C.~D.}\ \bibnamefont
  {Roberts}},\ }\href {\doibase 10.1103/PhysRevC.53.390} {\bibfield  {journal}
  {\bibinfo  {journal} {Phys. Rev. C}\ }\textbf {\bibinfo {volume} {53}},\
  \bibinfo {pages} {390} (\bibinfo {year} {1996})}\BibitemShut {NoStop}%
\bibitem [{\citenamefont {Alkofer}\ \emph {et~al.}(2002)\citenamefont
  {Alkofer}, \citenamefont {Watson},\ and\ \citenamefont
  {Weigel}}]{Alkofer:2002bp}%
  \BibitemOpen
  \bibfield  {author} {\bibinfo {author} {\bibfnamefont {R.}~\bibnamefont
  {Alkofer}}, \bibinfo {author} {\bibfnamefont {P.}~\bibnamefont {Watson}}, \
  and\ \bibinfo {author} {\bibfnamefont {H.}~\bibnamefont {Weigel}},\ }\href
  {\doibase 10.1103/PhysRevD.65.094026} {\bibfield  {journal} {\bibinfo
  {journal} {Phys. Rev. D}\ }\textbf {\bibinfo {volume} {65}},\ \bibinfo
  {pages} {094026} (\bibinfo {year} {2002})}\BibitemShut {NoStop}%
\bibitem [{\citenamefont {Blank}\ \emph {et~al.}(2011)\citenamefont {Blank},
  \citenamefont {Krassnigg},\ and\ \citenamefont {Maas}}]{Blank:2010pa}%
  \BibitemOpen
  \bibfield  {author} {\bibinfo {author} {\bibfnamefont {M.}~\bibnamefont
  {Blank}}, \bibinfo {author} {\bibfnamefont {A.}~\bibnamefont {Krassnigg}}, \
  and\ \bibinfo {author} {\bibfnamefont {A.}~\bibnamefont {Maas}},\ }\href
  {\doibase 10.1103/PhysRevD.83.034020} {\bibfield  {journal} {\bibinfo
  {journal} {Phys. Rev. D}\ }\textbf {\bibinfo {volume} {83}},\ \bibinfo
  {pages} {034020} (\bibinfo {year} {2011})}\BibitemShut {NoStop}%
\bibitem [{\citenamefont {Fischer}\ and\ \citenamefont
  {Alkofer}(2003)}]{Fischer:2003rp}%
  \BibitemOpen
  \bibfield  {author} {\bibinfo {author} {\bibfnamefont {C.~S.}\ \bibnamefont
  {Fischer}}\ and\ \bibinfo {author} {\bibfnamefont {R.}~\bibnamefont
  {Alkofer}},\ }\href {\doibase 10.1103/PhysRevD.67.094020} {\bibfield
  {journal} {\bibinfo  {journal} {Phys. Rev. D}\ }\textbf {\bibinfo {volume}
  {67}},\ \bibinfo {pages} {094020} (\bibinfo {year} {2003})}\BibitemShut
  {NoStop}%
\bibitem [{\citenamefont {Alkofer}\ \emph {et~al.}(2008)\citenamefont
  {Alkofer}, \citenamefont {Fischer},\ and\ \citenamefont
  {Williams}}]{Alkofer:2008et}%
  \BibitemOpen
  \bibfield  {author} {\bibinfo {author} {\bibfnamefont {R.}~\bibnamefont
  {Alkofer}}, \bibinfo {author} {\bibfnamefont {C.~S.}\ \bibnamefont
  {Fischer}}, \ and\ \bibinfo {author} {\bibfnamefont {R.}~\bibnamefont
  {Williams}},\ }\href {\doibase 10.1140/epja/i2008-10646-x} {\bibfield
  {journal} {\bibinfo  {journal} {Eur. Phys. J. A}\ }\textbf {\bibinfo {volume}
  {38}},\ \bibinfo {pages} {53} (\bibinfo {year} {2008})}\BibitemShut {NoStop}%
\bibitem [{\citenamefont {Bhagwat}\ \emph {et~al.}(2003)\citenamefont
  {Bhagwat}, \citenamefont {Pichowsky}, \citenamefont {Roberts},\ and\
  \citenamefont {Tandy}}]{Bhagwat:2003vw}%
  \BibitemOpen
  \bibfield  {author} {\bibinfo {author} {\bibfnamefont {M.~S.}\ \bibnamefont
  {Bhagwat}}, \bibinfo {author} {\bibfnamefont {M.~A.}\ \bibnamefont
  {Pichowsky}}, \bibinfo {author} {\bibfnamefont {C.~D.}\ \bibnamefont
  {Roberts}}, \ and\ \bibinfo {author} {\bibfnamefont {P.~C.}\ \bibnamefont
  {Tandy}},\ }\href {\doibase 10.1103/PhysRevC.68.015203} {\bibfield  {journal}
  {\bibinfo  {journal} {Phys. Rev. C}\ }\textbf {\bibinfo {volume} {68}},\
  \bibinfo {pages} {015203} (\bibinfo {year} {2003})}\BibitemShut {NoStop}%
\bibitem [{\citenamefont {Krassnigg}\ and\ \citenamefont
  {Roberts}(2004{\natexlab{a}})}]{Krassnigg:2003dr}%
  \BibitemOpen
  \bibfield  {author} {\bibinfo {author} {\bibfnamefont {A.}~\bibnamefont
  {Krassnigg}}\ and\ \bibinfo {author} {\bibfnamefont {C.~D.}\ \bibnamefont
  {Roberts}},\ }\href {\doibase 10.1016/S0375-9474(04)00291-X} {\bibfield
  {journal} {\bibinfo  {journal} {Nucl. Phys. A}\ }\textbf {\bibinfo {volume}
  {737}},\ \bibinfo {pages} {7} (\bibinfo {year}
  {2004}{\natexlab{a}})}\BibitemShut {NoStop}%
\bibitem [{\citenamefont {Krassnigg}\ and\ \citenamefont
  {Roberts}(2004{\natexlab{b}})}]{Krassnigg:2003wy}%
  \BibitemOpen
  \bibfield  {author} {\bibinfo {author} {\bibfnamefont {A.}~\bibnamefont
  {Krassnigg}}\ and\ \bibinfo {author} {\bibfnamefont {C.~D.}\ \bibnamefont
  {Roberts}},\ }\href {http://fizika.phy.hr/fizika_b/bv04/b13p143.htm}
  {\bibfield  {journal} {\bibinfo  {journal} {Fizika B}\ }\textbf {\bibinfo
  {volume} {13}},\ \bibinfo {pages} {143} (\bibinfo {year}
  {2004}{\natexlab{b}})}\BibitemShut {NoStop}%
\bibitem [{\citenamefont {Fischer}\ and\ \citenamefont
  {Pennington}(2006)}]{Fischer:2005nf}%
  \BibitemOpen
  \bibfield  {author} {\bibinfo {author} {\bibfnamefont {C.~S.}\ \bibnamefont
  {Fischer}}\ and\ \bibinfo {author} {\bibfnamefont {M.~R.}\ \bibnamefont
  {Pennington}},\ }\href@noop {} {\bibfield  {journal} {\bibinfo  {journal}
  {Phys. Rev. D}\ }\textbf {\bibinfo {volume} {73}},\ \bibinfo {pages} {034029}
  (\bibinfo {year} {2006})}\BibitemShut {NoStop}%
\bibitem [{\citenamefont {Eichmann}\ \emph
  {et~al.}(2008{\natexlab{b}})\citenamefont {Eichmann}, \citenamefont
  {Krassnigg}, \citenamefont {Schwinzerl},\ and\ \citenamefont
  {Alkofer}}]{Eichmann:2008zz}%
  \BibitemOpen
  \bibfield  {author} {\bibinfo {author} {\bibfnamefont {G.}~\bibnamefont
  {Eichmann}}, \bibinfo {author} {\bibfnamefont {A.}~\bibnamefont {Krassnigg}},
  \bibinfo {author} {\bibfnamefont {M.}~\bibnamefont {Schwinzerl}}, \ and\
  \bibinfo {author} {\bibfnamefont {R.}~\bibnamefont {Alkofer}},\ }\href
  {\doibase 10.1016/j.ppnp.2007.12.018} {\bibfield  {journal} {\bibinfo
  {journal} {Prog. Part. Nucl. Phys.}\ }\textbf {\bibinfo {volume} {61}},\
  \bibinfo {pages} {84} (\bibinfo {year} {2008}{\natexlab{b}})}\BibitemShut
  {NoStop}%
\bibitem [{\citenamefont {Eichmann}\ \emph
  {et~al.}(2008{\natexlab{c}})\citenamefont {Eichmann}, \citenamefont
  {Alkofer}, \citenamefont {Cloet}, \citenamefont {Krassnigg},\ and\
  \citenamefont {Roberts}}]{Eichmann:2008ae}%
  \BibitemOpen
  \bibfield  {author} {\bibinfo {author} {\bibfnamefont {G.}~\bibnamefont
  {Eichmann}}, \bibinfo {author} {\bibfnamefont {R.}~\bibnamefont {Alkofer}},
  \bibinfo {author} {\bibfnamefont {I.~C.}\ \bibnamefont {Cloet}}, \bibinfo
  {author} {\bibfnamefont {A.}~\bibnamefont {Krassnigg}}, \ and\ \bibinfo
  {author} {\bibfnamefont {C.~D.}\ \bibnamefont {Roberts}},\ }\href {\doibase
  10.1103/PhysRevC.77.042202} {\bibfield  {journal} {\bibinfo  {journal} {Phys.
  Rev. C}\ }\textbf {\bibinfo {volume} {77}},\ \bibinfo {pages} {042202(R)}
  (\bibinfo {year} {2008}{\natexlab{c}})}\BibitemShut {NoStop}%
\bibitem [{\citenamefont {Eichmann}\ \emph {et~al.}(2009)\citenamefont
  {Eichmann}, \citenamefont {Cloet}, \citenamefont {Alkofer}, \citenamefont
  {Krassnigg},\ and\ \citenamefont {Roberts}}]{Eichmann:2008ef}%
  \BibitemOpen
  \bibfield  {author} {\bibinfo {author} {\bibfnamefont {G.}~\bibnamefont
  {Eichmann}}, \bibinfo {author} {\bibfnamefont {I.~C.}\ \bibnamefont {Cloet}},
  \bibinfo {author} {\bibfnamefont {R.}~\bibnamefont {Alkofer}}, \bibinfo
  {author} {\bibfnamefont {A.}~\bibnamefont {Krassnigg}}, \ and\ \bibinfo
  {author} {\bibfnamefont {C.~D.}\ \bibnamefont {Roberts}},\ }\href {\doibase
  10.1103/PhysRevC.79.012202} {\bibfield  {journal} {\bibinfo  {journal} {Phys.
  Rev. C}\ }\textbf {\bibinfo {volume} {79}},\ \bibinfo {pages} {012202(R)}
  (\bibinfo {year} {2009})}\BibitemShut {NoStop}%
\bibitem [{\citenamefont {Holl}\ \emph
  {et~al.}(2005{\natexlab{b}})\citenamefont {Holl}, \citenamefont {Krassnigg},
  \citenamefont {Roberts},\ and\ \citenamefont {Wright}}]{Holl:2004un}%
  \BibitemOpen
  \bibfield  {author} {\bibinfo {author} {\bibfnamefont {A.}~\bibnamefont
  {Holl}}, \bibinfo {author} {\bibfnamefont {A.}~\bibnamefont {Krassnigg}},
  \bibinfo {author} {\bibfnamefont {C.~D.}\ \bibnamefont {Roberts}}, \ and\
  \bibinfo {author} {\bibfnamefont {S.~V.}\ \bibnamefont {Wright}},\ }\href
  {\doibase doi:10.1142/S0217751X05023323} {\bibfield  {journal} {\bibinfo
  {journal} {Int. J. Mod. Phys. A}\ }\textbf {\bibinfo {volume} {20}},\
  \bibinfo {pages} {1778} (\bibinfo {year} {2005}{\natexlab{b}})}\BibitemShut
  {NoStop}%
\bibitem [{\citenamefont {Krassnigg}(2009)}]{Krassnigg:2009zh}%
  \BibitemOpen
  \bibfield  {author} {\bibinfo {author} {\bibfnamefont {A.}~\bibnamefont
  {Krassnigg}},\ }\href {\doibase 10.1103/PhysRevD.80.114010} {\bibfield
  {journal} {\bibinfo  {journal} {Phys. Rev. D}\ }\textbf {\bibinfo {volume}
  {80}},\ \bibinfo {pages} {114010} (\bibinfo {year} {2009})}\BibitemShut
  {NoStop}%
\bibitem [{\citenamefont {Blank}\ and\ \citenamefont
  {Krassnigg}(2011{\natexlab{c}})}]{Blank:2011ha}%
  \BibitemOpen
  \bibfield  {author} {\bibinfo {author} {\bibfnamefont {M.}~\bibnamefont
  {Blank}}\ and\ \bibinfo {author} {\bibfnamefont {A.}~\bibnamefont
  {Krassnigg}},\ }\href {\doibase 10.1103/PhysRevD.84.096014} {\bibfield
  {journal} {\bibinfo  {journal} {Phys. Rev. D}\ }\textbf {\bibinfo {volume}
  {84}},\ \bibinfo {pages} {096014} (\bibinfo {year}
  {2011}{\natexlab{c}})}\BibitemShut {NoStop}%
\bibitem [{\citenamefont {Popovici}\ \emph {et~al.}(2014)\citenamefont
  {Popovici}, \citenamefont {Hilger}, \citenamefont {Gomez-Rocha},\ and\
  \citenamefont {Krassnigg}}]{Popovici:2014pha}%
  \BibitemOpen
  \bibfield  {author} {\bibinfo {author} {\bibfnamefont {C.}~\bibnamefont
  {Popovici}}, \bibinfo {author} {\bibfnamefont {T.}~\bibnamefont {Hilger}},
  \bibinfo {author} {\bibfnamefont {M.}~\bibnamefont {Gomez-Rocha}}, \ and\
  \bibinfo {author} {\bibfnamefont {A.}~\bibnamefont {Krassnigg}},\ }\href@noop
  {} {\ }\Eprint {http://arxiv.org/abs/1407.7970} {1407.7970 [hep-ph]}
  \BibitemShut {NoStop}%
\bibitem [{\citenamefont {Olive}\ and\ \citenamefont {others (Particle
  Data~Group)}(2014)}]{Olive:2014rpp}%
  \BibitemOpen
  \bibfield  {author} {\bibinfo {author} {\bibfnamefont {K.~A.}\ \bibnamefont
  {Olive}}\ and\ \bibinfo {author} {\bibnamefont {others (Particle
  Data~Group)}},\ }\href@noop {} {\bibfield  {journal} {\bibinfo  {journal}
  {Chin. Phys.}\ }\textbf {\bibinfo {volume} {C38}},\ \bibinfo {pages} {090001}
  (\bibinfo {year} {2014})}\BibitemShut {NoStop}%
\bibitem [{\citenamefont {Brambilla}\ \emph {et~al.}(2010)\citenamefont
  {Brambilla} \emph {et~al.}}]{Brambilla:2010cs}%
  \BibitemOpen
  \bibfield  {author} {\bibinfo {author} {\bibfnamefont {N.}~\bibnamefont
  {Brambilla}} \emph {et~al.},\ }\href {\doibase
  10.1140/epjc/s10052-010-1534-9} {\bibfield  {journal} {\bibinfo  {journal}
  {Eur. Phys. J.}\ }\textbf {\bibinfo {volume} {C71}},\ \bibinfo {pages} {1534}
  (\bibinfo {year} {2010})}\BibitemShut {NoStop}%
\bibitem [{\citenamefont {Dorkin}\ \emph {et~al.}(2014)\citenamefont {Dorkin},
  \citenamefont {Kaptari}, \citenamefont {Hilger},\ and\ \citenamefont
  {Kampfer}}]{Dorkin:2013rsa}%
  \BibitemOpen
  \bibfield  {author} {\bibinfo {author} {\bibfnamefont {S.~M.}\ \bibnamefont
  {Dorkin}}, \bibinfo {author} {\bibfnamefont {L.~P.}\ \bibnamefont {Kaptari}},
  \bibinfo {author} {\bibfnamefont {T.}~\bibnamefont {Hilger}}, \ and\ \bibinfo
  {author} {\bibfnamefont {B.}~\bibnamefont {Kampfer}},\ }\href {\doibase
  10.1103/PhysRevC.89.034005} {\bibfield  {journal} {\bibinfo  {journal}
  {Phys.Rev.}\ }\textbf {\bibinfo {volume} {C89}},\ \bibinfo {pages} {034005}
  (\bibinfo {year} {2014})}\BibitemShut {NoStop}%
\end{thebibliography}

%

\end{document}